\documentclass[12pt,a4paper]{article}
\usepackage{epsfig}
\usepackage{mathtools}
\usepackage{amsfonts}
\usepackage{amssymb}
\usepackage{amsmath}
\usepackage{color}
\usepackage{slashed}
\usepackage{cite}
\usepackage{caption}
\usepackage{subcaption}

\usepackage{tikz}
\usetikzlibrary{shapes}
\usetikzlibrary{trees}
\usetikzlibrary{calc}
\usetikzlibrary{decorations.pathmorphing}
\usetikzlibrary{decorations.markings}

\def\ep              {\varepsilon }

\def\la              {\langle}
\def\ra              {\rangle}

\def\ZZ              {\mathcal{Z}}

\newcommand{\abr}[1]{\langle #1 \rangle}
\newcommand{\sbr}[1]{\left[ #1 \right]}

\newcommand{\vll}{{\smash{\lambda}}}
\newcommand{\vlt}{{\smash{\tilde{\lambda}}}}
\newcommand{\vlet}{{\smash{\tilde{\eta}}}}
\newcommand{\vle}{{\smash{\eta}}}

\newcommand{\vlluu}{\smash{\underline{\underline{\smash{\lambda}}}}}
\newcommand{\vltuu}{\smash{\underline{\underline{\smash{\tilde{\lambda}}}}}}
\newcommand{\vleuu}{\smash{\underline{\underline{\smash{\tilde{\eta}}}}}}

\definecolor{grayn}{gray}{0.7}
\definecolor{lightgrayn}{gray}{0.8}

\newlength{\vacuumradius}
\newlength{\onshellradius}
\tikzstyle{db}=[circle, black, fill=black, minimum width=\onshellradius, draw, inner sep=0pt]
\tikzstyle{dw}=[circle, black, fill=white, minimum width=\onshellradius, draw, inner sep=0pt]
\tikzstyle{dvac}=[circle, black, fill=lightgrayn, minimum width=\vacuumradius, inner sep=0pt]
\tikzstyle{dl}=[circle, black, fill=white, inner sep=2pt]

\tikzset{
    gluon/.style={decorate, decoration={coil, amplitude=1.2pt, segment length=3pt, aspect=1}, draw=black}
}

\tikzset{->-/.style={decoration={
			markings,
			mark=at position .5 with {\arrow{>}}},postaction={decorate}}}

\def\d{\text{d}}
\def\e{\text{e}}

\begin{document}

\begin{center}

\vspace{1cm}

{\bf \Large The all-loop conjecture for  integrands of \\ reggeon amplitudes in $\mathcal{N}=4$ SYM} \vspace{1cm}

{\large A.E. Bolshov$^{1,2}$, L.V. Bork$^{3,4,5}$, A.I. Onishchenko$^{1,2,6}$}\vspace{0.5cm}

{\it
	$^1$Bogoliubov Laboratory of Theoretical Physics, Joint
		Institute for Nuclear Research, Dubna, Russia, \\
	$^2$Moscow Institute of Physics and Technology (State University), Dolgoprudny, Russia, \\
	$^3$Institute for Theoretical and Experimental Physics, Moscow,
	Russia,\\
	$^4$The Center for Fundamental and Applied Research, All-Russia
	Research Institute of Automatics, Moscow, Russia, \\
	$^5$National Research Nuclear University (MEPhI), Moscow,
	Russia,\\
	$^6$Skobeltsyn Institute of Nuclear Physics, Moscow State University, Moscow, Russia}\vspace{1cm}

\abstract{In this paper we present the all-loop conjecture for integrands of Wilson line form factors, also known as reggeon amplitudes, in $\mathcal{N}=4$ SYM. In particular we present a new gluing operation in momentum twistor space used to obtain reggeon tree-level amplitudes and loop integrands starting from corresponding expressions for on-shell amplitudes. The introduced gluing procedure is used to derive the BCFW recursions both for tree-level reggeon amplitudes and their loop integrands. In addition we provide predictions for the reggeon loop integrands written in the basis of local integrals. As a check of the correctness of the gluing operation at loop level we derive the expression for LO BFKL kernel in $\mathcal{N}=4$ SYM.}
\end{center}

\begin{center}
\end{center}

\newpage

\tableofcontents{}\vspace{0.5cm}

\renewcommand{\theequation}{\thesection.\arabic{equation}}

\section{Introduction}

In the last two decades the tremendous progress in understanding of the
structure of S-matrix (amplitudes) in gauge theories in various dimensions has been achieved. The most prominent examples of such progress are various results for the amplitudes in
$\mathcal{N}=4$ SYM theory. See for review \cite{Henrietta_Amplitudes,Talesof1001Gluons}. These results were near to impossible to obtain without plethora of new ideas and approaches to the perturbative computations in gauge theories. These new ideas and approaches mostly exploit analytical properties of amplitudes rather then rely on standard textbook Feynman diagram technique.

It is important to note that these
analyticity based approaches appear to be effective not only for computations of the amplitudes but for form factors and correlation functions of local and non-local operators in $\mathcal{N}=4$ SYM and other gauge theories as well \cite{FormFactorsPeriodicWilsonLoops,HarmonyofFF_Brandhuber,Engelund:2012re,PolytopesFormFactors,SuperFormFactorsHalfBPS,FormFactorsGrassmanians,SoftTheoremsFormFactors,TwistorFormFactors1,TwistorFormFactors2,TwistorFormFactors3}. So for many important results for the amplitudes in $\mathcal{N}=4$ SYM their analogs for the form factors and correlation functions were found \cite{FormFactorsPeriodicWilsonLoops,HarmonyofFF_Brandhuber,GeneralizedUnitarityFormFactors,PolytopesFormFactors,FormFactorsGrassmanians,SoftTheoremsFormFactors,TwistorFormFactors1,TwistorFormFactors2,TwistorFormFactors3,BrandhuberConnectedPrescription,HeConnectedFormulaFormFactors,DilatationOperatorFormFactors5,FormFactorsYsystem1,FormFactorsYsystem2,offshell-1leg,offshell-multiple}. First of all, new variables such as helicity spinors and momentum twistors appear also useful for the description of the form factors and correlation functions
\cite{FormFactorsPeriodicWilsonLoops,HarmonyofFF_Brandhuber,TwistorFormFactors1}. At tree level various recurrence relations (BCFW, CSW e t.c.) were constructed for the form factors of some
local \cite{FormFactorsPeriodicWilsonLoops,HarmonyofFF_Brandhuber,BKV_SuperForm,PolytopesFormFactors,SuperFormFactorsHalfBPS} as well as non-local \cite{KotkoWilsonLines} operators and various closed solutions for such recurrence relations were obtained \cite{FormFactorsPeriodicWilsonLoops,HarmonyofFF_Brandhuber,PolytopesFormFactors,SuperFormFactorsHalfBPS,offshell-1leg,offshell-multiple,q2zeroFormFactors,TwistorFormFactors1,TwistorFormFactors2,TwistorFormFactors3}. Ultimately for the form factors of operators from stress tensor supermultiplet \cite{FormFactorsGrassmanians} as well as for Wilson line operators \cite{offshell-1leg} the representation in terms of integral over Grassmannian was discovered\footnote{So
in this sense these objects at tree level are known for arbitrary number of external legs.}. Also in the case of the Wilson line operators such representation was generalized to the form factors with arbitrary number of Wilson line operator insertions as well as correlation functions \cite{offshell-multiple}. Dual description for such objects in terms of twistor string theories was investigated and in this context different CHY like representations for form factors were obtained \cite{BrandhuberConnectedPrescription,HeConnectedFormulaFormFactors,ambitwistorFormfactors}. In addition, unconventional (compared to standard textbooks) geometrical interpretation for the correlation functions of stress tensor supermultiplet operators was conjectured \cite{Eden:2017fow} similar to the ``Amplituhedron'' \cite{Amplituhdron_1,Amplituhdron_2,Amplituhdron_3,Amplituhdron_7,Amplituhdron_8} for the amplitudes.
At loop level various other results were obtained for the form factors at high orders of perturbation theory and/or number of external particles \cite{FormFactorsPeriodicWilsonLoops,Bork2010FormFact,GeneralizedUnitarityFormFactors,2loopFormFactors2012,DilatationOperatorFormFactors5,DilatationOperatorFormFactors1,DilatationOperatorFormFactors2,DilatationOperatorFormFactors3,DilatationOperatorFormFactors4,FormFactorsColorKinematic,FormFactorsColorKinematic5loop,HennGehrmann_FormF3loops_2011,Boels:2015yna,Banerjee:2016kri} and connection between form factors and integrable systems \cite{DilatationOperatorFormFactors6,FormFactorsGrassmanians} was discussed. Interesting
results \cite{Huang2016FormFactorBoundaryContribution,Engelund:2015cfa} also should be mentioned.

The ultimate goal for such investigations, similar to the amplitude case, would be the evaluation, in some closed form, of all factors and correlation functions off all possible operators in  $\mathcal{N}=4$ SYM at arbitrary value of coupling constant.

In this note we are going to continue to work in this direction and consider the possibility of constructing recurrence relations for the loop integrands of the Wilson line form factors in $\mathcal{N}=4$ SYM theory.

Wilson lines are non-local gauge invariant operators and are interesting objects not only from pure theoretical but also from phenomenological point of view. They appear, for example, in the description of reggeon amplitudes  in the framework of Lipatov's effective QCD lagrangian\footnote{We also want to mention recent work \cite{Kotko:2017nkx}, where Wilson lines arise in the process of off-shell analytic continuation of light-front quantized Yang-Mills action.}   \cite{vanHamerenBCFW1,vanHamerenBCFW2,vanHamerenBCFW3,LipatovEL1,LipatovEL2,KirschnerEL1,KirschnerEL2,
KotkoWilsonLines,vanHamerenWL1,vanHamerenWL2,vanHameren:2017txa,vanHameren:2017hxx}, within the context of
$k_T$  or high-energy factorization
\cite{GribovLevinRyskin,CataniCiafaloniHautmann,CollinsEllis,CataniHautmann} as well as in the study of processes at multi-Regge kinematics. The Wilson line operators play the role of sources for the reggeized gluons, while their form factors are directly related to amplitudes with reggeized gluons in such framework. The results in this field to a large extent originate from long lasting efforts of St.Peterburg and Novosibirsk groups in the investigation of asymptotic behavior of QFT scattering amplitudes at high energies (Regge limit),  which can be tracked back in time to early works \cite{GribovReggeonDiagrams} of Gribov. These results, in particular, include  resummation of leading high energy logarithms $(\alpha_s\ln s)^n$ to all orders in strong coupling constant (LLA resummation) in QCD, which eventually resulted in the discovery of Balitsky-Fadin-Kuraev-Lipatov (BFKL) equation \cite{BFKL1,BFKL2,BFKL3,BFKL4,BFKL5} governing the LLA  high energy asymptotic behavior of QCD scattering amplitudes. Today BFKL equation is known at next-to-leading-logarithmic-approximation (NLLA) \cite{NLOBFKL1,NLOBFKL2,NLOreggeization}. The current
article can be also considered as an effort in this direction, namely towards NNLLA BFKL in the context of $\mathcal{N}=4$ SYM. More accurately, the results of this article can be considered as a solution to the problem of the reduction of individual Feynman diagrams to a set of master integrals in BFKL computations. As in general amplitudes multiloop calculations, in BFKL calculations there are basically two steps in getting the final result.
The first one is the reduction of contributing individual Feynman diagrams to a finite set of so called
master integrals and the second one is the evaluation of these master integrals themselves. In this paper
we discuss only the first part of this problem, which is the easiest one.

This article is organized as follows. In section 2 we remind the reader the definition of the Wilson line form factors and correlation functions as well as give the definition of so called gluing operator $\hat{A}_{i-1i}$ considered in our previous papers \cite{ambitwistorFormfactors}. This operator allows one to convert the on-shell amplitudes into the Wilson line form factors and will be heavily used throughout the paper. In section 3 we discuss how the BCFW recurrence relations for the Wilson line form factors are constructed with the use of helicity spinor variables used to describe kinematical data. After that we show how the mentioned BCFW recursion can be derived from the BCFW recursion for on-shell amplitudes by means of application of the gluing operators. Section 4 contains the derivation of the gluing operator in the case when kinematical data are encoded by momentum twistor variables. In section 5 we remind the reader the necessary facts about BCFW recursion for integrands of the on-shell amplitudes in momentum twistor space. After that we show how applying the gluing operator one can formulate similar recurrence relation for the Wilson line form factor integrands as well. We also show how one can directly transform the integrands of on-shell amplitudes into the integrands of the Wilson line form factors on the examples of local on-shell integrands. After that we perform simple but interesting self consistency check of our considerations. Namely starting from our results for tree and one loop level Wilson line form factors we correctly reproduce LO BFKL kernel. Appendix \ref{appA} contains the derivation of the Grassmannian integral representation for the reggeon amplitudes starting from corresponding representation for the on-shell amplitudes.

\section{Form factors of Wilson lines and gluing operation}
\label{ReggeonsWilsonLines}

\subsection{Form factors of Wilson lines operators}

To describe the form factors of Wilson line operators
we will use the definition in \cite{KotkoWilsonLines}:
\begin{eqnarray}\label{WilsonLineOperDef}
\mathcal{W}_p^c(k) = \int d^4 x e^{ix\cdot k} \mathrm{Tr} \left\{
\frac{1}{\pi g} t^c \; \mathcal{P} \exp\left[\frac{ig}{\sqrt{2}}\int_{-\infty}^{\infty}
ds \; p\cdot A_b (x+ sp) t^b\right]
\right\}\, ,
\end{eqnarray}
where $t^c$ is $SU(N_c)$ generator\footnote{The color generators are normalized as $\mathrm{Tr} (t^a t^b) = \delta^{a b}$}, $k$ ($k^2 \neq 0$) is the off-shell reggeized gluon momentum and $p$ is its direction or polarization vector, such that $p^2=0$ and $p\cdot k = 0$. The polarization vector and momentum of the reggeized gluon are related to each other through the so called  $k_T$ - decomposition of the latter:
\begin{eqnarray}\label{kT}
k^{\mu} = x p^{\mu} + k_T^{\mu} \, ,\quad x \in [0,1] \, .
\end{eqnarray}
It is convenient to parametrize such decomposition by an auxiliary light-cone four-vector $q^{\mu}$, so that
\begin{eqnarray}
k_T^{\mu} (q) = k^{\mu} - x(q) p^{\mu}\quad \text{with}\quad x(q) = \frac{q\cdot k}{q\cdot p} \;\; \text{and} \;\; q^2 = 0.
\end{eqnarray}
Noting that the transverse momentum $k_T^{\mu}$ is orthogonal to both $p^\mu$ and $q^\mu$ vectors, we may write down the latter in the basis of two ``polarization'' vectors\footnote{Here we used the helicity spinor \cite{Henrietta_Amplitudes} decomposition of light-like four-vectors $p$ and $q$. We will also sometime abuse spinor helicity formalism notations and write
$\la q|\gamma^{\mu}|p]/2\equiv |p]\la q|$, $\lambda_q\equiv\la q|$ and
$\tilde{\lambda}_q\equiv[p|$.} \cite{vanHamerenBCFW1}:
\begin{eqnarray}
k_T^{\mu} (q) = -\frac{\kappa}{2}\frac{\la p|\gamma^{\mu}|q]}{[pq]}
- \frac{\kappa^{*}}{2}\frac{\la q|\gamma^{\mu}|p]}{\la qp\ra}\quad
\text{with} \quad \kappa = \frac{\la q|\slashed{k}|p]}{\la qp\ra},\;
\kappa^{*} = \frac{\la p|\slashed{k}|q]}{[pq]}.
\end{eqnarray}
It is easy to check, that $k^2 = -\kappa\kappa^{*}$ and both $\kappa$ and $\kappa^{*}$ variables are independent of auxiliary four-vector $q^{\mu}$ \cite{vanHamerenBCFW1}. Also, it turns out convenient to use spinor helicity decomposition of the light-cone four-vector $q$  as $q = |\xi\ra [\xi|$. $\mathcal{W}_p^c(k)$ is non-local gauge invariant operator and plays the role of source for the reggeized gluon \cite{BalitskyLectures,KotkoWilsonLines}, so
the form factors of such operators, or off-shell gauge invariant scattering amplitudes in our terminology, are closely related to the reggeon scattering amplitudes, and we will use words off-shell gauge invariant scattering amplitudes, reggeon amplitudes and Wilson line form factors hereafter as synonyms.

Both usual and color ordered reggeon amplitudes with $n$ reggeized and $m$ usual on-shell gluons could be then written in terms of the form factors with multiple Wilson line insertions as \cite{KotkoWilsonLines}:
\begin{eqnarray}\label{AmplitudeSeveralOffShellGluons}
A_{m+n}^* \left(1^{\pm},\ldots ,m^{\pm},g_{m+1}^*,\ldots ,g_{n+m}^*\right) =
\la \{k_i, \epsilon_i, c_i\}_{i=1}^m |\prod_{j=1}^n\mathcal{W}_{p_{m+j}}^{c_{m+j}}(k_{m+j})|0\ra\, .
\end{eqnarray}
Here asterisk denotes an off-shell gluon and $p$, $k$, $c$ are its direction, momentum and color index. Next $\la \{k_i, \epsilon_i, c_i\}_{i=1}^m|=\bigotimes_{i=1}^m\la k_i,\varepsilon_i, c_i|$ and $\la k_i,\varepsilon_i, c_i|$ denotes an on-shell gluon state  with momentum $k_i$, polarization vector $\varepsilon_i^-$ or $\varepsilon_i^+$ and color index $c_i$,
$p_j$ is the direction of the $j$'th ($j=1,...,n$) off-shell gluon and $k_j$ is its off-shell momentum. To simplify things, here we are dealing with color ordered amplitudes only. The usual amplitudes are then obtained using their color decomposition, see \cite{offshell-1leg,DixonReview}. For example, the color ordered amplitude with one reggeon and two on-shell gluons with opposite helicity at tree level is given by the following expression:
\begin{eqnarray}\label{Component3poingOffhsellAmpl}
A_{2+1}^*(1^-,2^+,g^*_3)
 &=&\frac{\delta^4(\vll_{1}\vlt_{1}+\vll_{2}\vlt_{2}+k_3)}{\kappa_3^*} \frac{\abr{p_3\, 1}^4}
 {\abr{p_3\, 1}\abr{1\, 2}\abr{2\, p_3}}.
\end{eqnarray}

When dealing with $\mathcal{N}=4$ SYM we may also consider other on-shell states from $\mathcal{N}=4$ supermultiplet. The easiest way to do it is to consider color ordered superamplitudes defined on $\mathcal{N}=4$ on-shell momentum superspace \cite{DrummondSuperconformalSymmetry,Nair}:
\begin{eqnarray}\label{AmplitudeSeveralOffShellGluonsSUSY}
A_{m+n}^* \left(\Omega_1,\ldots,\Omega_m,g_{m+1}^*,\ldots ,g_{n+m}^*\right) =
\la \Omega_1\ldots\Omega_m|
\prod_{j=1}^n\mathcal{W}_{p_{m+j}}(k_{m+j})|0\ra,
\end{eqnarray}
where $\la \Omega_1\Omega_2\ldots\Omega_m|\equiv \bigotimes_{i=1}^m\la 0|\Omega_i$ and $\Omega_i$ ($i=1,...,m$) denotes an $\mathcal{N}=4$ on-shell chiral superfield \cite{Nair}:
\begin{eqnarray}
\Omega= g^+ + \vlet_A\psi^A + \frac{1}{2!}\vlet_{A}\vlet_{B}\phi^{AB}
+ \frac{1}{3!}\vlet_A\vlet_B\vlet_C\epsilon^{ABCD}\bar{\psi}_{D}
+ \frac{1}{4!}\vlet_A\vlet_B\vlet_C\vlet_D\epsilon^{ABCD}g^{-},
\end{eqnarray}
Here, $g^+, g^-$ are creation/annihilation operators of gluons with $+1$ and $-1$ helicities, $\psi^A$, $\bar{\psi}_A$ stand for creation/annihilation operators of four Weyl spinors with negative helicity $-1/2$ and four Weyl spinors with  positive helicity correspondingly, while $\phi^{AB}$ denote creation/annihilation operators for six scalars (anti-symmetric in the $SU(4)_R$ $R$-symmetry indices $AB$).
The $A_{m+n}^* \left(\Omega_1,\ldots,g_{n+m}^*\right)$ superamplitude is then the function of the following kinematic\footnote{We used helicity spinor decomposition of on-shell particles momenta.} and Grassmann variables
\begin{eqnarray}\label{AmplitudeSeveralOffShellGluonsArgumentsSUSY}
A_{k,m+n}^* \left(\Omega_1,\ldots,g_{m+n}^*\right) =A_{k,m+n}^*\left(\{\lambda_i,\tilde{\lambda}_i,\tilde{\eta}_i\}_{i=1}^{m};
\{k_i,\lambda_{p,i},\tilde{\lambda}_{p,i}\}_{i=m+1}^{m+n}\right).
\end{eqnarray}
and encodes in addition to the amplitudes with gluons also amplitudes with other on-shell states similar to the case of usual on-shell superamplitudes \cite{Henrietta_Amplitudes}. Here, additional index\footnote{We hope there will be no confusion with momentum labels.} $k$ in $A_{k,m+n}^*$ denotes the total degree of $A_{k,m+n}^*$ in Grassmann variables $\eta_i$, which is given by $4k-4n$. For example the supersymmetrised (in on-shell states) version of
(\ref{Component3poingOffhsellAmpl}) is given by:
\begin{eqnarray}\label{Super3poingOffhsellAmpl}
 A_{2,2+1}^*(\Omega_1, \Omega_{2},g_3^*) &=& \prod_{A=1}^4\frac{\partial}{\partial\vlet_{p_3}^A} \left[\frac{\delta^4(\vll_{1}\vlt_{1}+\vll_{2}\vlt_{2}+k_3)}{\kappa_3^*} \frac{
 	\delta^8(\vll_{p_3}\vlet_{p_3}+\vll_{1}\vlet_{1}+\vll_{2}\vlet_{2})
 }{\abr{p_3\, 1}\abr{1\, 2}\abr{2\, p_3}} \right] \nonumber \\
 &=&\frac{\delta^4(\vll_{1}\vlt_{1}+\vll_{2}\vlt_{2}+k_3)}{\kappa_3^*} \frac{\delta^4\left(\tilde{\eta}_{1}\abr{p_3\, 1}+\tilde{\eta}_{2}\abr{p_3\, 2}\right)}
 {\abr{p_3\, 1}\abr{1\, 2}\abr{2\, p_3}}.
 \end{eqnarray}
Here we have $k=2$, $m=2$ and $n=1$. We also for simplicity will often drop $\partial^4/\partial\vlet^4_{{p_i}}$ projectors in further considerations.

\subsection{Gluing operator: transforming on-shell amplitudes into Wilson line form factors}

In \cite{ambitwistorReggeons,ambitwistorFormfactors} it was conjectured that one can compute the form factors of Wilson line operators by means of the four dimensional ambitwistor string theory \cite{ambitwistorString4d}. In an addition to the standard vertex operators ${\mathcal V}$ and $\widetilde{\mathcal{V}}$, which describe $\Omega_i$ on-shell states in $\mathcal{N}=4$ SYM field theory, one can introduce, so called,
generalised vertex operators $\mathcal{V}^{gen.}$ \cite{ambitwistorFormfactors}:
\begin{eqnarray}
\mathcal{V}^{gen.}_j\sim\int A^*_{2,2+1}(\Omega_j,\Omega_{j+1},g^*)
\prod_{i=j,j+1}\mathcal{V}_i
~\frac{ \d^2\vll_i \d^2\vlt_i}{\text{Vol[GL(1)]}} ~\d^4\vlet_i.
\end{eqnarray}
Then it was conjectured that the following relation holds at least at tree level:
\begin{eqnarray}\label{stringCorrelator}
	A_{k,m+n}^* \left(\Omega_1,\ldots,g_{m+n}^*\right)=
	\langle {\mathcal V}_1,\ldots {\mathcal V}_m \mathcal{V}^{gen.}_{m+1},\ldots,\mathcal{V}^{gen.}_{m+n}\rangle_{worldsheet~fields}.
\end{eqnarray}
Here $\la \ldots \ra$ means average with respect to string worldsheet fields. This conjecture was successfully verified at the level of Grassmannian integral representations for the whole tree level S-matrix \cite{ambitwistorReggeons,ambitwistorFormfactors} and on several particular examples \cite{ambitwistorFormfactors} with fixed number of external states. Effectively
the evaluation of the string theory correlation function in (\ref{stringCorrelator}) can
be reduced to the action of some integral operator $\hat{A}$ on the on-shell amplitudes.
In the case of one Wilson line operator insertion the relation between on-shell amplitude
and the Wilson line form factor looks like:
\begin{equation}
A_{n+1}^* = \hat{A}_{n+1,n+2}[A_{n+2}]\, ,
\end{equation}
where $A_{n+2}$ is the usual on-shell superamplitude with $n+2$ on-shell external states and the gluing integral operator $\hat{A}_{n+1,n+2}$ acts on the kinematical variables associated with the states $\Omega_{n+1}$ and $\Omega_{n+2}$.

The action of $\hat{A}_{n+1,n+2}$ on any function $f$ of variables $\left\{
\vll_i , \vlt_i , \vlet_i
\right\}_{i=1}^{n+2}$ is formally given by
\begin{equation}
\hat{A}_{n+1,n+2}[f] \equiv \int\prod_{i=n+1}^{n+2}\frac{d^2\vll_i d^2\vlt_i d^4\vlet_i}{\text{Vol}[GL(1)]} A_{2,2+1}^* (g^*, \Omega_{n+1}, \Omega_{n+2}) \times f\left(
\left\{
\vll_i , \vlt_i , \vlet_i
\right\}_{i=1}^{n+2}
\right). \label{gluingdef}
\end{equation}
This expression can be simplified.
Performing integration over  $\vlt_{n+1}$, $\vlt_{n+2}$, $\vlet_{n+1}$ and $\vlet_{n+2}$ variables \cite{ambitwistorFormfactors} in (\ref{gluingdef}) we get
\begin{eqnarray}
\hat{A}_{n+1,n+2}[f]=\frac{\langle p_{n+1} \xi_{n+1}\rangle}{\kappa_{n+1}^*}\int \frac{d\beta_1}{\beta_1}\wedge\frac{d\beta_2}{\beta_2}~\frac{1}{\beta_1^2\beta_2} ~ f\left(\{\lambda_i,\tilde{\lambda}_i,\tilde{\eta}_i\}_{i=1}^{n+2}\right)\big{|}_{*}, \label{Aoperator}
\end{eqnarray}
where $\big{|}_{*}$ denotes substitutions $\{\lambda_i,\tilde{\lambda}_i,\eta_i\}_{i=n+1}^{n+2}
\mapsto\{\lambda_i(\beta),\tilde{\lambda}_i(\beta),\tilde{\eta}_i(\beta)\}_{i=n+1}^{n+2}$ with
\begin{align}
&\vll_{n+1}(\beta) = \vlluu_{n+1} + \beta_2\vlluu_{n+2}\, , && \vlt_{n+1}(\beta) =
\beta_1\vltuu_{n+1}  + \frac{(1+\beta_1)}{\beta_2}\vltuu_{n+2}\, ,
&&\vlet_{n+1}(\beta) = -\beta_1\vleuu_{n+1}\, , \nonumber\\
&\vll_{n+2}(\beta) = \vlluu_{n+2} + \frac{(1+\beta_1)}{\beta_1\beta_2}\vlluu_{n+1}\, ,
&& \vlt_{n+2}(\beta) = -\beta_1\vltuu_{n+2} -\beta_1\beta_2 \vltuu_{n+1}\, , &&\vlet_{n+2}(\beta) = \beta_1\beta_2\vleuu_{n+1}\, .
\end{align}\label{starSubst1}
and
\begin{eqnarray}
\vlluu_{n+1}=\lambda_p,~\vltuu_{n+1}=\frac{\la\xi |k}{\la\xi p\ra},~\vleuu_{n}=\vlet_p;
~~\vlluu_{n+2}=\lambda_{\xi},~\vltuu_{n+2}=\frac{\la p |k}{\la\xi p\ra},~\vleuu_{n+2}=0.
\end{eqnarray}
All other variables left unshifted.

The integration with respect to $\beta_{1,2}$ will be understood as
a residue form \cite{DualitySMatrix} and will be evaluated by means of the composite residue in points $res_{\beta_2=0}\circ res_{\beta_1=-1}$.
For example, one can obtain \cite{ambitwistorFormfactors} the Wilson line form factor $A^*_{3,3+1}(1^-,2^+,3^-,g_4^*)$ from 5 point NMHV on-shell amplitude $A^*_{3,5}(1^-,2^+,3^-,4^-,5^+)$:
\begin{eqnarray}\label{Exampl1}
A^*_{3,3+1}(1^-,2^+,3^-,g_4^*)=\hat{A}_{45}[A^*_{3,5}(1^-,2^+,3^-,4^-,5^+)],
\end{eqnarray}
where \cite{vanHamerenBCFW1,offshell-1leg}
\begin{eqnarray}
A^*_{3,3+1}(1^-,2^+,3^-,g_4^*)=\delta^{4}\left(\sum_{i=1}^3\lambda_i\tilde{\lambda}_i+k_4\right)\frac{1}{\kappa_4}\frac{[2p_4]^4}{[12][23][3p_4][p_41]}.
\end{eqnarray}

Several Wilson line operator insertions correspond to the consecutive action of
several gluing operators. For example $A^*_{3,0+3}(g_1^*,g_2^*,g_3^*)$ can be obtained
 \cite{ambitwistorFormfactors}
from 6 point NMHV amplitude $A_{3,6}(1^-2^+3^-4^+5^-6^+)$:
\begin{equation}\label{Exampl2}
A^*_{3,0+3}(g_1^*,g_2^*,g_3^*)
=(\hat{A}_{12}\circ\hat{A}_{34}\circ\hat{A}_{56})[A_{3,6}(1^-2^+3^-4^+5^-6^+)],
\end{equation}
where $A^*_{3,0+3}$ is given by ($\mathbb{P}'$ is the permutation operator which shifts all spinor and momenta labels by +1 mod 3.):
\begin{eqnarray}\label{3reggeonAmplitude}
A^*_{3,0+3}(g^*_1,g^*_2,g^*_3)&=&\delta^4(k_1+k_2+k_3)
\left(1+\mathbb{P}'+\mathbb{P}'^2\right)f,\nonumber\\
f&=&\frac{\langle
	p_1p_2\rangle^3[p_2p_3]^3}{\kappa_3\kappa^*_1\langle p_2|k_1|p_3]\langle
	p_1|k_3|p_2]\langle p_2|k_1|p_2]}.
\end{eqnarray}

There is another way of representing the action of gluing operator. One can note
that (\ref{gluingdef}) is in fact equivalent to the action of a pair of consecutive
BCFW bridge operators, in terminology of \cite{AmplitudesPositiveGrassmannian}, on the $f$ function weighted with an inverse
soft factor. Namely, if one \cite{AmplitudesPositiveGrassmannian} defines
$[i,j\rangle$ BCFW shift operator as $Br(i,j)$
which acts on the function $f$ of the arguments $\left\{
\vll_i , \vlt_i , \vlet_i
\right\}_{i=1}^{n+2}$, $1 \leq i,j \leq n+2$
according to:
\begin{eqnarray}
&&Br(i,i+1)\left[f(\ldots,\lambda_i,\tilde{\lambda}_i\tilde{\eta}_i,\ldots,\lambda_j,\tilde{\lambda}_j,\tilde{\eta}_j,\ldots)\right]=\int \frac{\d\alpha }{\alpha}f(\ldots,\lambda_i,\hat{\tilde{\lambda}}_i,\hat{\tilde{\eta}}_i,\ldots,\hat{\lambda}_j,\tilde{\lambda}_j,\tilde{\eta}_j,\ldots)=\nonumber\\
&&\int \frac{\d\alpha }{\alpha}f(\ldots,\lambda_i,
\tilde{\lambda}_i-\alpha\tilde{\lambda}_j,
\tilde{\eta}_i-\alpha\tilde{\eta}_j,\ldots,
\lambda_j+\alpha\lambda_i,
\tilde{\lambda}_j,\tilde{\eta}_j,\ldots),
\end{eqnarray}
then one can see that the following relation holds:
\begin{equation}\label{GluibgOperAsBCFWBridges}
\hat{A}_{n+1,n+2}[~f~]=Br(n+1,n+2)\circ Br(n+2,n+1)\left[S^{-1}(1,n+2,n+1)~f~\right],
\end{equation}
where
\begin{eqnarray}\label{SoftFactor}
S(1,n+2,n+1)=\frac{\kappa_{n+1}^*\langle 1 n+1\rangle}
{\langle 1 n+2\rangle\langle n+2 n+1\rangle},
\end{eqnarray}
and function the $f$ depends on $\{\vll_i , \vlt_i , \vlet_i \}_{i=1}^{n+2}$ arguments.
\begin{figure}[t]
	\begin{center}
		\epsfxsize=5cm
		\epsffile{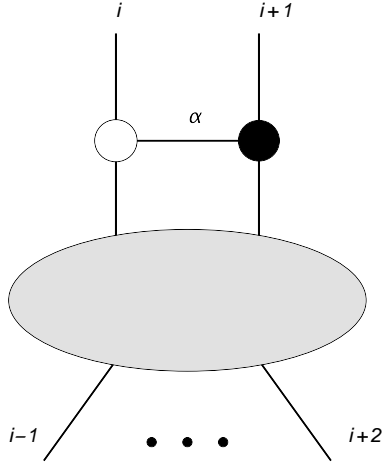}
	\end{center}\vspace{-0.2cm}
	\caption{The action of $Br(i,i+1)$ operator on the on-shell diagram. White blob is $\overline{\mbox{MHV}}_3$ amplitude, black one - $\mbox{MHV}_3$.}\label{BCFW_Br_Oper}
\end{figure}

Note also that since $Br(i,j)$ operators act naturally on on-shell diagrams \cite{AmplitudesPositiveGrassmannian}
one can easily consider the action of $\hat{A}_{n+1,n+2}$ operator on
the top-cell diagram corresponding to the $A_{k,n+2}$ tree level on-shell amplitude.
The top-cell for $A_{k,n+2}$ in its turn can be represented as the integral over Grassmannian
$L_{n+2}^k$ \cite{AmplitudesPositiveGrassmannian} (here let's ignore integration contour for a moment):
\begin{eqnarray}
L_{n+2}^k = \int \frac{d^{k\times n+2} C}{\text{Vol}[GL(k)]}
\frac{\delta^{k\times 2} (C\cdot\vlt)\delta^{k\times 4} (C\cdot\vlet)\delta^{(n+2-k)\times 2} (C^{\perp}\cdot\vll)}{(1\cdots k)(2\cdots k+1)\cdots (n+2\cdots k-1)}.
\end{eqnarray}
Then one can see that the following relation also holds:
\begin{eqnarray}\label{GluingGrassmannianIntegrals}
\hat{A}_{n+1,n+2}\left[L_{n+2}^k\right]=\Omega^k_{n+1}\, ,
\end{eqnarray}
where $\Omega^k_{n+2}$ is the Grassmannian integral representation
for the off-shell amplitude $A_{k,n+1}^*$, with the Wilson line insertion positioned
after the on-shell state with number $n$ \cite{offshell-1leg},
if the appropriate integration contour is chosen for $\Omega^k_{n+2}$\footnote{One can think of this as alternative derivation of the results of Appendix \ref{appA} of\cite{ambitwistorFormfactors}. See also Appendix of the current article for notation explanation.}:
\begin{eqnarray}\label{GrIntWLFF}
\Omega_{n+2}^k  =
\int\frac{d^{k\times (n+2)}C'}{\text{Vol}[GL (k)]}Reg.
\frac{\delta^{k\times 2} \left( C'
	\cdot \vltuu \right)
	\delta^{k\times 4} \left(C' \cdot \vleuu \right)
	\delta^{(n+2-k)\times 2} \left(C'^{\perp} \cdot \vlluu \right)}{(1 \cdots k)\cdots (n+1 \cdots k-2) (n+2 \; 1\cdots k-1)},
\end{eqnarray}
with
\begin{equation}\label{RegFunction}
Reg.=\frac{\la\xi_{n+1} p_{n+1}\ra}{\kappa_{n+1}^{*}}\frac{(n+2 \; 1\cdots k-1)}{(n+1 \; 1\cdots k-1)}.
\end{equation}
and
\begin{align}\label{SpinorsInDeformadGrassmannian}
&\vlluu_i = \vll_i, & i = 1,\ldots  n& , &\vlluu_{n+1} &= \lambda_{p_{n+1}} ,
&\vlluu_{n+2} &= \xi_{n+1} \nonumber \\
&\vltuu_i = \vlt_i, & i = 1,\ldots  n& ,
&\vltuu_{n+1} &= \frac{\la\xi_{n+1} |k_{n+1}}{\la\xi_{n+1}\, p_{n+1}\ra},
&\vltuu_{n+2} &= - \frac{\la p_{n+1} |k_{n+1}}{\la\xi_{n+1}\, p_{n+1}\ra} , \nonumber \\
&\vleuu_i = \vlet_i,  & i = 1,\ldots  n& , &\vleuu_{n+1} &= \vlet_{p_{n+1}} ,
&\vleuu_{n+2} &= 0 . \nonumber \\
\end{align}
The action of several $\hat{A}_{i,i+1}$ operators can be considered among the same lines
and the result reproduces Grassmannian representation of the form factors with multiple Wilson line operator insertion obtained in \cite{offshell-multiple}.

At the end of this section let us make the following comment. Both on-shell and off-shell amplitudes (Wilson line form factors) can be represented by means of the BCFW recursion relations. But due to different analytical properties (Wilson line form factors will have additional type of poles corresponding to the Wilson line propagators \cite{vanHamerenBCFW1})
the recursion for on-shell and off-shell amplitudes looks rather different.
However, from the examples similar to ones considered above (namely (\ref{Exampl1}) and (\ref{Exampl1})) one can note that not only gluing operator maps on-shell amplitudes to off-shell ones but one can choose
representation for the on-shell amplitude in terms of the BCFW recursion in such a way that each BCFW term from on-shell amplitude will be mapped one-to-one to the terms from the BCFW recursion for the off-shell amplitudes. So a natural question to ask is whether it is possible to \emph{derive} the BCFW recursion for the Wilson line form factors from the BCFW recursion for the on-shell
amplitudes. We will address this question in the next section.

\section{BCFW recursion for Wilson line form factors}
\label{offshellBCFW}

\subsection{Off-shell BCFW from analyticity}

First let us remind the reader the main results of \cite{vanHamerenBCFW1} and comment on supersymmetric extension of the off-shell BCFW recursion. The off-shell BCFW recursion for the reggeon amplitudes with an arbitrary number of off-shell reggeized gluons was worked out in \cite{vanHamerenBCFW1}.
Similar to the BCFW recursion \cite{BCFW1,BCFW2} for the on-shell amplitudes it is based on the observation, that a contour integral of an analytical function $f$ vanishing at infinity equals to zero, that is
\begin{eqnarray}
\oint\frac{d z}{2\pi i}\frac{f (z)}{z} = 0 .
\end{eqnarray}
and the integration contour expands to infinity. Taking the above integral by residues we get
\begin{eqnarray}
f (0) = - \sum_i \frac{\text{res}_i f(z)}{z_i} , \label{fpolessum}
\end{eqnarray}
where the sum is over all poles of $f$ and $\text{res}_i f(z)$ is a residue of $f$ at pole $z_i$. Using this, one can relate the off-shell amplitude to the sum over contributions of its factorisation
channels, which in turn can be represented as the off-shell amplitudes with smaller number of external states. In the original on-shell BCFW recursion the $z$-dependence of scattering amplitude is obtained by a $z$-dependent shift of particle's momenta. Similarly, the off-shell gluon BCFW recursion of \cite{vanHamerenBCFW1} is formulated using a shift of momenta for two external gluons $i$ and $j$ with a vector

\begin{eqnarray}
e^{\mu} = \frac{1}{2}\la p_i|\gamma^{\mu}|p_j] ,\qquad p_i\cdot e = p_j\cdot e = e\cdot e = 0 ,
\end{eqnarray}
so that
\begin{align}
\hat{k}_i^{\mu}(z) &\equiv  k_i^{\mu} + z e^{\mu} = x_i (p_j) p_i^{\mu}
- \frac{\kappa_i - [p_i p_j] z}{2}\frac{\la p_i|\gamma^{\mu} |p_j]}{[p_i p_j]}
- \frac{\kappa_i^{*}}{2}\frac{\la p_j|\gamma^{\mu}|p_i]}{\la p_j p_i\ra}\; , \\
\hat{k}_j^{\mu}(z) &\equiv k_j^{\mu} - z e^{\mu} = x_j (p_i) p_j^{\mu}
- \frac{\kappa_j}{2}\frac{\la p_j|\gamma^{\mu} |p_i]}{[p_j p_i]}
- \frac{\kappa_j^{*} + \la p_i p_j\ra z}{2}\frac{\la p_i|\gamma^{\mu} |p_j]}{\la p_i p_j\ra} .
\end{align}
This shift does not violate momentum conservation and we still have $p_i\cdot \hat{k}_i (z) = 0$ and $p_j\cdot\hat{k}_j (z) = 0$. We would like to note, that the overall effect of shifting momenta is that the values of $\kappa_i$ and $\kappa_j^{*}$ shift, while $\kappa_i^{*}$ and $\kappa_j$ stay unshifted. In the on-shell limit the above shift corresponds to the usual $[i, j\ra$ BCFW shift. Note also, that we could have chosen another shift vector $e^{\mu} = \frac{1}{2}\la p_j|\gamma^{\mu}| p_i]$ and shift $\kappa_i^{*}$ and $\kappa_j$ instead. The off-shell amplitudes we consider in this paper do also have a correct large $z$ ($z\to\infty$) behavior \cite{vanHamerenBCFW1}, so that we should not worry about boundary terms at infinity.

The sum over the poles (\ref{fpolessum}) for $z$-dependent off-shell gluon scattering amplitude is given by the following graphical representation\footnote{We are considering the color ordered scattering amplitudes and without loss of generality may use shift of two adjacent legs $1$ and $n$.} \cite{vanHamerenBCFW1}:
\begin{eqnarray} \label{graphicalBCFW1}
\begin{tikzpicture}[baseline={($(nc.base) - (0,0)$)},transform shape, scale=0.6]
\node (nc) at (0,0) {};
\node[below] (n1) at (-1.5,-1) {1};
\node[below] (nn) at (1.5,-1) {$n$};
\node[left] (n2) at (-1.5,0.2) {2};
\node[right] (nnm1) at (1.5,0.2) {$n-1$};
\node[right] (d1) at (-1,1) {};
\node[above] (d2) at (0,1.2) {};
\node[left] (d3) at (1,1) {};
\draw[double, thick] (nc) -- (n1);
\draw[double, thick] (nc) -- (nn);
\draw[double, thick] (nc) -- (n2);
\draw[double, thick] (nc) -- (nnm1);
\draw[very thick] (nc) circle [radius=0.6];
\draw[fill,grayn] (nc) circle [radius=0.6];
\draw[fill] (d1) circle [radius=0.05];
\draw[fill] (d2) circle [radius=0.05];
\draw[fill] (d3) circle [radius=0.05];
\end{tikzpicture}
= \sum_{i=2}^{n-2}\sum_{h} \mathbb{A}_{i,h} + \sum_{i=2}^{n-1}\mathbb{B}_i + \mathbb{C} + \mathbb{D},
\end{eqnarray}
where
\begin{gather}
\mathbb{A}_{i,h} =
\begin{tikzpicture}[baseline={($(nc.base) - (0,0.1)$)},transform shape, scale=0.6]
\node (nc) at (0,0) {};
\coordinate (ch) at (1,0.01);
\node[below] (n1) at (-1,-1.5) {$\hat{1}$};
\node[above] (ni) at (-1,1.5) {$i$};
\node[above] (nh) at (1,0) {$h$};
\node[left] (d1) at (-0.8,-0.5) {};
\node[left] (d2) at (-1,0) {};
\node[left] (d3) at (-0.8,0.5) {};
\draw[double, thick] (nc) -- (n1);
\draw[double, thick] (nc) -- (ni);
\draw[ultra thick] (nc) -- (ch);
\draw[very thick] (nc) circle [radius=0.6];
\draw[fill,grayn] (nc) circle [radius=0.6];
\draw[fill] (d1) circle [radius=0.05];
\draw[fill] (d2) circle [radius=0.05];
\draw[fill] (d3) circle [radius=0.05];
\end{tikzpicture}
\; \frac{1}{k_{1,i}^2} \;
\begin{tikzpicture}[baseline={($(nc.base) - (0,0.1)$)},transform shape, scale=0.6]
\node (nc) at (0,0) {};
\coordinate (ch) at (-1,0.01);
\node[below] (nn) at (1,-1.5) {$\hat{n}$};
\node[above] (nip1) at (1,1.5) {$i+1$};
\node[above] (nh) at (-1,0) {$-h$};
\node[right] (d1) at (0.8,-0.5) {};
\node[right] (d2) at (1,0) {};
\node[right] (d3) at (0.8,0.5) {};
\draw[double, thick] (nc) -- (nn);
\draw[double, thick] (nc) -- (nip1);
\draw[ultra thick] (nc) -- (ch);
\draw[very thick] (nc) circle [radius=0.6];
\draw[fill,grayn] (nc) circle [radius=0.6];
\draw[fill] (d1) circle [radius=0.05];
\draw[fill] (d2) circle [radius=0.05];
\draw[fill] (d3) circle [radius=0.05];
\end{tikzpicture}
\qquad
\mathbb{B}_i =
\begin{tikzpicture}[baseline={($(nc.base) - (0,0.1)$)},transform shape, scale=0.6]
\node (nc) at (0,0) {};
\node[below] (n1) at (-1,-1.5) {$\hat{1}$};
\node[above] (nim1) at (-1,1.5) {$i-1$};
\node[right] (nisynm) at (0,1.9) {$i$};
\node[above] (ni) at (0,1.7) {};
\node (nh) at (1.5,0) {};
\node[left] (d1) at (-0.8,-0.5) {};
\node[left] (d2) at (-1,0) {};
\node[left] (d3) at (-0.8,0.5) {};
\draw[gluon] (nc) -- (ni);
\draw[thick] (nc) -- (ni);
\draw[gluon] (nc) -- (nh);
\draw[thick] (nc) -- (nh);
\draw[double, thick] (nc) -- (n1);
\draw[double, thick] (nc) -- (nim1);
\draw[very thick] (nc) circle [radius=0.6];
\draw[fill,grayn] (nc) circle [radius=0.6];
\draw[fill] (d1) circle [radius=0.05];
\draw[fill] (d2) circle [radius=0.05];
\draw[fill] (d3) circle [radius=0.05];
\end{tikzpicture}
\;\; \frac{1}{2 p_i\cdot k_{i,n}} \;\;
\begin{tikzpicture}[baseline={($(nc.base) - (0,0.1)$)},transform shape, scale=0.6]
\node (nc) at (0,0) {};
\node[below] (nn) at (1,-1.5) {$\hat{n}$};
\node[above] (nip1) at (1,1.5) {$i+1$};
\node[left] (nisynm) at (0,1.9) {$i$};
\node[above] (ni) at (0,1.7) {};
\node (nh) at (-1.5,0) {};
\node[right] (d1) at (0.8,-0.5) {};
\node[right] (d2) at (1,0) {};
\node[right] (d3) at (0.8,0.5) {};
\draw[gluon] (nc) -- (ni);
\draw[thick] (nc) -- (ni);
\draw[gluon] (nc) -- (nh);
\draw[thick] (nc) -- (nh);
\draw[double, thick] (nc) -- (nn);
\draw[double, thick] (nc) -- (nip1);
\draw[very thick] (nc) circle [radius=0.6];
\draw[fill,grayn] (nc) circle [radius=0.6];
\draw[fill] (d1) circle [radius=0.05];
\draw[fill] (d2) circle [radius=0.05];
\draw[fill] (d3) circle [radius=0.05];
\end{tikzpicture} \nonumber \\ \nonumber \\
\mathbb{C} = \frac{1}{\kappa_1} \;
\begin{tikzpicture}[baseline={($(nc.base) - (0,0)$)},transform shape, scale=0.6]
\node (nc) at (0,0) {};
\node[below] (n1) at (-1.5,-1) {$\hat{1}$};
\node[below] (nn) at (1.5,-1) {$\hat{n}$};
\node[left] (n2) at (-1.5,0.2) {2};
\node[right] (nnm1) at (1.5,0.2) {$n-1$};
\node[right] (d1) at (-1,1) {};
\node[above] (d2) at (0,1.2) {};
\node[left] (d3) at (1,1) {};
\draw[ultra thick] (nc) -- (n1);
\draw[double, thick] (nc) -- (nn);
\draw[double, thick] (nc) -- (n2);
\draw[double, thick] (nc) -- (nnm1);
\draw[very thick] (nc) circle [radius=0.6];
\draw[fill,grayn] (nc) circle [radius=0.6];
\draw[fill] (d1) circle [radius=0.05];
\draw[fill] (d2) circle [radius=0.05];
\draw[fill] (d3) circle [radius=0.05];
\end{tikzpicture}
\qquad
\mathbb{D} = \frac{1}{\kappa_n^{*}} \;
\begin{tikzpicture}[baseline={($(nc.base) - (0,0)$)},transform shape, scale=0.6]
\node (nc) at (0,0) {};
\node[below] (n1) at (-1.5,-1) {$\hat{1}$};
\node[below] (nn) at (1.5,-1) {$\hat{n}$};
\node[left] (n2) at (-1.5,0.2) {2};
\node[right] (nnm1) at (1.5,0.2) {$n-1$};
\node[right] (d1) at (-1,1) {};
\node[above] (d2) at (0,1.2) {};
\node[left] (d3) at (1,1) {};
\draw[double, thick] (nc) -- (n1);
\draw[ultra thick] (nc) -- (nn);
\draw[double, thick] (nc) -- (n2);
\draw[double, thick] (nc) -- (nnm1);
\draw[very thick] (nc) circle [radius=0.6];
\draw[fill,grayn] (nc) circle [radius=0.6];
\draw[fill] (d1) circle [radius=0.05];
\draw[fill] (d2) circle [radius=0.05];
\draw[fill] (d3) circle [radius=0.05];
\end{tikzpicture} \label{graphicalBCFW2}
\end{gather}
$k_{i,j}^{\mu}\equiv k_i^{\mu} + k_{i+1}^{\mu} + \cdots + k_j^{\mu}$ and $h$ is an internal on-shell gluon helicity or a summation index over all on-shell states in the Nair on-shell supermultiplet in the supersymmetric case discussed later. Here and below we use the convention that double lines may stand both for off-shell and on-shell gluons.
The coil crossed with a line correspond to the off-shell gluons (Wilson line operator insertion).
The thick solid lines stand for on-shell particles. The off-shell coil lines can be bent apart to form a single eikonal quark lines \cite{KotkoWilsonLines,vanHamerenBCFW1}.  According to this $k_j^{\mu}$ in $k_{i,j}^{\mu}$ can be
either off-shell or on-shell depending on the context.

Let's now discuss each type of the terms encountered in (\ref{graphicalBCFW1}) in more details.
The $\mathbb{A}_{i,h}$ terms are usual on-shell BCFW terms, which correspond to the $z$ - poles at which denominator of internal gluon (and also fermion or scalar) propagator $\hat{k}_{1,i}^2 (z)$ vanishes:
\begin{eqnarray}\label{OnShellConditionProp}
	\hat{k}_{1,i}^2 (z) = 0.
\end{eqnarray}
This is standard BCFW on-shell condition for physical states of $\mathcal{N}=4$ SYM supermultiplet.

The $\mathbb{B}_i$ term is a new one and is unique to the BCFW recursion for the off-shell amplitudes. It originates from the situation when the denominators of eikonal propagators coming from Wilson line expansion vanish, that is
\begin{eqnarray}\label{EikonalOnShellCondition}
	p_i\cdot\hat{k}_{i,n} (z) = 0
\end{eqnarray}
and $p_i^{\mu}$ is the direction of the Wilson line associated with the off-shell gluon. It is important to understand that condition
$p_i\cdot\hat{k}_{i,n} (z) = 0$ fixes only the \emph{direction}  of momentum flowing through the Wilson line $\hat{k}_{i,n}$. The off-shell momenta $\hat{k}_{L}=k_{i-1}+\ldots+\hat{k}_{1}$ and $\hat{k}_{R}=k_{i}+\ldots+\hat{k}_{n}$,
which belongs to the off-shell amplitudes in Fig. \ref{graphicalBCFW2}, term $\mathbb{B}_i$, are different ($\hat{k}_{L} = - \hat{k}_{R}$), but satisfy the same condition
$p_i\cdot\hat{k}_{L/R}= 0$. We also want to stress that this term is present only if $i$ labels an off-shell external gluon. In addition, let us note that $\kappa^*_i$ factors in the pair of off-shell amplitudes, which contribute to this term,
are given explicitly by the following expressions:
\begin{eqnarray}
	\kappa_{i,L}^*=\frac{\langle p_i|\hat{k}_{1}+\ldots+k_{i-1}|q_i]}{[p_iq_i]},~
	\kappa_{i,R}^*=\frac{\langle p_i|\hat{k}_{n}+\ldots+k_{i+1}|q_i]}{[p_iq_i]},
\end{eqnarray}
where $\kappa_{i,L}^*$ and $\kappa_{i,R}^*$ belong to the off-shell amplitudes
positioned to the left and to the right in Fig. \ref{graphicalBCFW2} for the term $\mathbb{B}_i$.

The $\mathbb{C}$ term is only present if the gluon number $1$ is off-shell. It is also unique to the BCFW recursion for the off-shell amplitudes. It appears due to vanishing of the external momentum square
\begin{eqnarray}
	\hat{k}_1^2 (z)=0.
\end{eqnarray}
Similarly, the $\mathbb{D}$ term is due to vanishing of the external momentum square $\hat{k}_n^2 (z)$. It turns out that both these contributions could be calculated in terms of the same BCFW term with the off-shell gluons $1$ or $n$ exchanged for the on-shell ones. The helicity of the on-shell gluons depends on the type
of the term ($\mathbb{C}$ or $\mathbb{D}$) and the shift vector $e^{\mu}$ ($\frac{1}{2}\la p_i|\gamma^{\mu}|p_j]$ or $\frac{1}{2}\la p_j|\gamma^{\mu}|p_i]$) used. We refer the reader to \cite{vanHamerenBCFW1} for further details and examples.

The use of shifts involving only on-shell legs also allows one to perform the supersymmetrization of the off-shell BCFW recursion introduced in \cite{vanHamerenBCFW1}. Indeed, it is easy to see, that the supersymmetric shifts of momenta and corresponding Grassmann variables are given by the on-shell BCFW $[i, j\ra$ super-shifts\footnote{These shifts respect both momentum and supermomentum conservation.}:
\begin{eqnarray}
|\hat{i}] = |1] + z|j], \qquad |\hat{j}\ra = |j\ra - z|i\ra, \qquad \hat{\eta}_A^i = \eta_A^i + z\eta_A^j.
\end{eqnarray}
No other spinors or Grassmann variables shift.

\subsection{Off-shell BCFW from gluing operation}\label{BCFWfromGluing}
\begin{figure}[t]
	\begin{center}
		\epsfxsize=13cm
		\epsffile{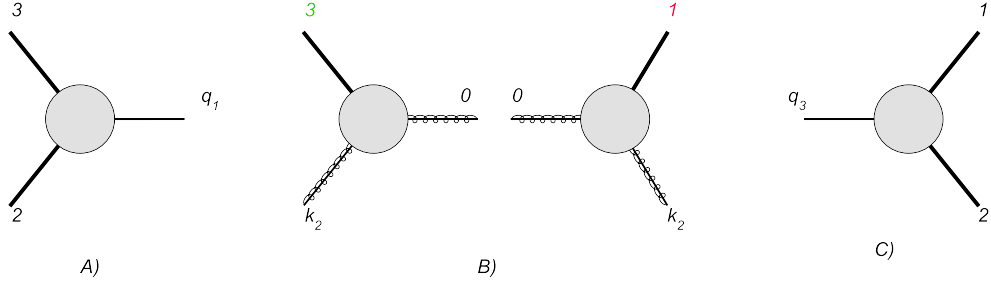}
	\end{center}\vspace{-0.2cm}
	\caption{The off-shell BCFW contributions representing $A^*_{0+3}(g_1^*,g_2^*,g_3^*)$ for
	shift of the $k_1$ and $k_3$ external momenta. The first and the third terms labeled A) and
	C) are type $\mathbb{C}$ and $\mathbb{D}$ contributions, while the second, labeled B), term is type $\mathbb{B}$ contribution of the off-shell BCFW recursion (\ref{graphicalBCFW1}).}\label{BCFW_A36}
\end{figure}
\begin{figure}[t]
	\begin{center}
		\epsfxsize=18cm
		\epsffile{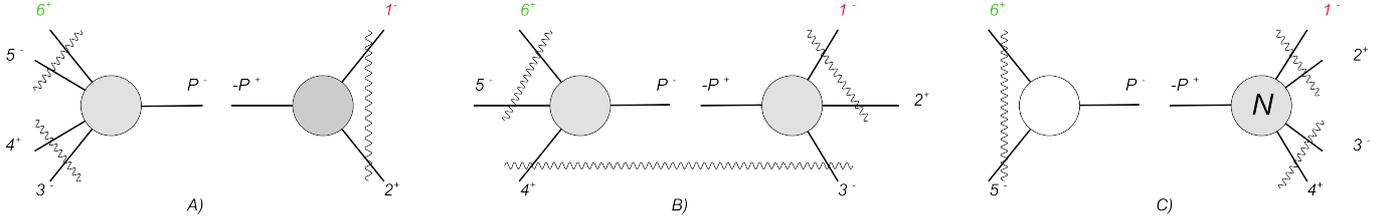}
	\end{center}\vspace{-0.2cm}\hspace{-2cm}
	\caption{The action of the gluing operators $\hat{A}_{ii+1}$ (represented by wavy line) on $[\textcolor{red}{1}^-,\textcolor{green}{6}^+ \rangle$ BCFW representation of $A_{3,6}(1^-,2^+,3^-,4^+,5^-,6^+)$ amplitude.}\label{BCFW6point}
\end{figure}

The aim of this section is to derive the off-shell recursion relations described above from the BCFW recursion for the on-shell amplitudes by means of the gluing operator. Before proceeding with general derivation let us consider a simple example first: we will take the BCFW recursion
for the on-shell 6-point NMHV amplitude $A_{3,6}(1^-,2^+,3^-,4^+,5^-,6^+)$ and transform it
into the three point off-shell amplitude $A^{*}_{0+3}(g_1^*,g_2^*,g^*_3)$ considered in \cite{vanHamerenBCFW1}. This off-shell amplitude in its turn also can be obtained from the off-shell BCFW recursion, when
external momenta 1 and 3 are shifted. Contributions corresponding to this shift are given in Fig. \ref{BCFW_A36}, and the sum of these three terms is given by (\ref{3reggeonAmplitude}).

So let's consider $A_{3,6}(1^-,2^+,3^-,4^+,5^-,6^+)$ amplitude represented via the standard
BCFW shift $[\textcolor{red}{1}^-,\textcolor{green}{6}^+ \rangle$.
The amplitude is then given, once again,
by three terms (see Fig. \ref{BCFW6point})
\begin{eqnarray}
A_{3,6}(1^-,2^+,3^-,4^+,5^-,6^+)=A+B+C.
\end{eqnarray}
where
\begin{eqnarray}
A&=&A_{2,5}(3^-,4^+,5^-,\hat{6}^+,\hat{P}^+)\frac{1}{q_{1,2}^2}
A_{2,3}(\hat{1}^-,2^+,-\hat{P}^-),\\
B&=&A_{2,4}(4^+,5^-,\hat{6}^+,\hat{P}^-)\frac{1}{q_{1,3}^2}
A_{2,4}(\hat{1}^-,2^+,3^-,-\hat{P}^+),\\
C&=&A_{1,3}(5^-,\hat{6}^+,\hat{P}^+)\frac{1}{q_{5,6}^2}
A_{3,5}(\hat{1}^-,2^+,3^-,4^+,-\hat{P}^-).
\end{eqnarray}
Here $q_{a,b} = \sum_{i=a}^b q_i$ and $q_i$ denote on-shell particle momenta.

Now we are going to consider the action of our gluing operators on $A_{3,6}$ on-shell amplitude, which will
convert all pairs of $i^-,(i+1)^+$ gluons into Wilson line operator insertions
(reggeized gluons).  As we have discussed in previous chapters to do this we have to take into account the action of the following combination of the gluing operations on $A_{3,6}$:
\begin{eqnarray}
A_{0+3}(g^*_1,g^*_2,g^*_3)=\hat{A}_{12}\circ \hat{A}_{34}\circ \hat{A}_{56}[A_{3,6}(1^-,2^+,3^-,4^+,5^-,6^+)].
\end{eqnarray}
Let's consider each contribution in details.

We will start with $A$ contribution first:
\begin{eqnarray}
\hat{A}_{12}\circ \hat{A}_{34}\circ \hat{A}_{56}[A]=
\hat{A}_{34}\circ \hat{A}_{56}[A_{2,5}(\hat{6}^+,5^-,4^+,3^-,\hat{P}^-)]
\hat{A}_{12}\left[\frac{1}{p_{1,2}^2}A_{2,3}(\hat{1}^-,2^+,-\hat{P}^+)\right].\nonumber\\
\end{eqnarray}
It turns out, that both the value of the BCFW shift parameter $z$ as
well as shifted spinors are regular after we made $\big{|}_*$ substitutions (see (\ref{Aoperator}) and (\ref{starSubst1})) corresponding to gluing operations and took limits $\beta_2\rightarrow 0,~\beta_1\rightarrow -1$. Let us introduce the following notation for the spinors entering $k_T$ - decomposition of momenta of three reggeized gluons, each spinor will be labeled by corresponding gluing operator:
\begin{eqnarray}
\hat{A}_{12}~\mapsto~|p_1\rangle,~|\xi_1\rangle;~
\hat{A}_{34}~\mapsto~|p_2\rangle,~|\xi_2\rangle;~
\hat{A}_{56}~\mapsto~|p_3\rangle,~|\xi_3\rangle.
\end{eqnarray}
The original value of the on-shell $z$ parameter ($z=\frac{[21]}{[62]}$) transformed
after the action of $\hat{A}_{12}$ into
\begin{equation}
z = \frac{[21]}{[62]} \mapsto \frac{\kappa_1 \abr{\xi_3\, p_3}}{\kappa_3^* \sbr{p_3\, p_1}}
\end{equation}
and helicity spinor decomposition of momentum $\hat{P}$ is given now by
\begin{equation}
|\hat{P}\rangle =  x (p_3)|p_1\rangle - \frac{\kappa_1^*}{\abr{p_3\, p_1}}|p_3\rangle\, ,\quad |\hat{P}] = |p_1]\, ,
\end{equation}
where we used   $k_T$ - decomposition of the first reggeized gluon $g_1^*$  momentum $k_1$:
\begin{equation}
k_1 = x (p_3) p_1 - \frac{\kappa_1}{\sbr{p_1\, p_3}}|p_1\ra [p_3| - \frac{\kappa_1^*}{\abr{p_3\, p_1}}|p_3\ra [p_1|
\end{equation}
Then, it is easy to see that
\begin{eqnarray}
\hat{A}_{34}\circ \hat{A}_{56}[A_{2,5}(3^-, 4^+, 5^-, \hat{6}^+,\hat{P}^+)]=
A_{1+2}^*(g_2^*,\hat{g}_3^*,\hat{P}^+),
\end{eqnarray}
where $\hat{g}_3^*$ denotes reggeized gluon $g_3^*$ with momentum shifted as
\begin{equation}
\hat{k}_3 = k_3 + \frac{\kappa_1}{\sbr{p_1\, p_3}} |p_1\ra [p_3 |.
\end{equation}
For the term
\begin{eqnarray}
\hat{A}_{12}\left[\frac{1}{q_{1,2}^2}A_{2,3}(\hat{1}^-,2^+,-\hat{P}^-)\right]
\end{eqnarray}
we have
\begin{eqnarray}
\hat{A}_{12}\left[\frac{1}{q_{1,2}^2}A_{2,3}(\hat{1}^-,2^+,-\hat{P}^-)\right]
=res_{\beta_1=-1}\circ res_{\beta_2=0}[\omega_A],
\end{eqnarray}
where
\begin{eqnarray}
\omega_A&=& -\frac{\abr{p_1\, \xi_1}}{\kappa^*_1}~\left(\frac{1}{k_1^2}~
\frac{\langle 1\hat{P}\rangle^3}{\langle 12\rangle\langle 2\hat{P}\rangle}\right)\Big{|}_*
~\frac{1}{\beta_1^2\beta_2}
\frac{d\beta_1\wedge d\beta_2}{\beta_1\beta_2}\nonumber\\
&=& \frac{1}{\kappa_1}\frac{1}{(1+\beta_1)}\frac{d\beta_1\wedge d\beta_2}{\beta_1\beta_2} + \mbox{less singular terms}.
\end{eqnarray}
Taking residues and combining everything together we finally get for
$A$ term
\begin{eqnarray}\label{Dterm}
\hat{A}_{12}\circ\hat{A}_{34}\circ\hat{A}_{56}[A]
=\frac{1}{\kappa_1}A_{1+2}^*(\hat{g}_3^*,g_2^*,\hat{P}^+),
\end{eqnarray}
which is precisely the $\mathbb{C}$ term from the off-shell BCFW recursion \cite{vanHamerenBCFW1} for $A^*_{0+3}(g_1^*,g_2^*,g_3^*)$ reggeon amplitude (see (\ref{graphicalBCFW1})). The $C$ term can be analysed similarly. In this case we get
\begin{eqnarray}
\hat{A}_{12}\circ\hat{A}_{34}\circ\hat{A}_{56}[C]
=\frac{1}{\kappa_3}A_{1+2}^*(\hat{g}_1^*,g_2^*,\hat{P}^-)\, ,
\end{eqnarray}
where helicity decomposition of momentum $\hat{P}$ is given by
\begin{eqnarray}
|\hat{P}]=\left(x(p_1)|p_3]-
\frac{\kappa_1}{[p_3\, p_1]}|p_1]\right),\quad
|\hat{P}\rangle=|p_3\rangle.
\end{eqnarray}
This is precisely the $\mathbb{D}$ term from the off-shell BCFW recursion \cite{vanHamerenBCFW1} for
$A^*_{0+3}(g_1^*,g_2^*,g_3^*)$ amplitude.

\begin{figure}[t]
	\begin{center}
		\epsfxsize=17cm
		\epsffile{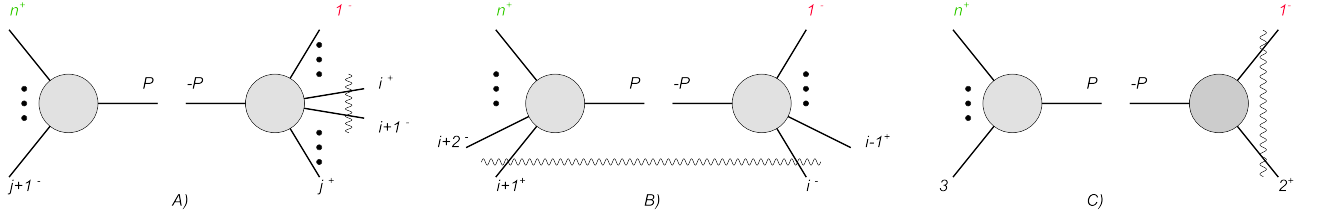}
	\end{center}\vspace{-0.2cm}
	\caption{The action of the gluing operators $\hat{A}_{ii+1}$ on the individual BCFW terms of the on-shell $[\textcolor{red}{1}^-,\textcolor{green}{n}^+ \rangle$ recurcion in general case. $A)$ diagrams will give $\mathbb{A}$ type terms of the off-shell BCFW recurcion, $B)$ type diagrams will give $\mathbb{B}$ type terms, while $C)$ diagrams will give $\mathbb{C}$ and $\mathbb{D}$ type terms of the off-shell BCFW recurcion.}\label{generalBCFW}
\end{figure}
Now let us turn to $B$ contribution to $A_{3,6}$ (see Fig. \ref{BCFW_A36}). The value of $z$ parameter in this case transforms under the action of the gluing operator as
\begin{equation}
z = \frac{q_{1,3}^2}{\la 1| q_2 + q_3| 6]} \mapsto \frac{\la p_2| k_1+k_2 |p_2]}{\abr{p_1\, p_2}\sbr{p_2\, p_3}}.
\end{equation}
Here it is convenient to consider first the action of $\hat{A}_{34}$. In this case the value of $\hat{P}$ momentum is given by
\begin{equation}
\hat{P} = \hat{q}_1 + q_2 + \frac{(1+\beta_1)}{\beta_2}\kappa_2^* \frac{|p_2\ra [p_2|}{\abr{\xi_2\, p_2}} + \mbox{less singular terms},
\end{equation}
before residues evaluation.
So for the whole $B$ term after $\hat{A}_{34}$ action we have
\begin{eqnarray}
\hat{A}_{34}[B]&=&\hat{A}_{34}\left[A_{2,4}(4^+, 5^-, \hat{6}^+,\hat{P}^-)\frac{1}{q_{1,3}^2}
A_{2,4}(\hat{1}^-,2^+,3^-,-\hat{P}^+)\right]\nonumber\\
&=&res_{\beta_1=-1}\circ res_{\beta_2=0}[\omega_B]
\end{eqnarray}
where
\begin{multline}
\omega_B = \frac{\abr{p_2\, \xi_2}}{\kappa_2^*}\left(
\frac{\abr{5\, \hat{P}}^4}{\abr{4\, 5}\abr{5\, \hat{6}}\abr{\hat{6}\, \hat{P}}\abr{\hat{P}\, 4}} \frac{1}{q_{1,3}^2} \frac{\abr{\hat{1}\, 3}^4}{\abr{\hat{1}\, 2}\abr{2\, 3}\abr{3\, \hat{P}}\abr{\hat{P}\, 1}}
\right)\Big|_*~\frac{1}{\beta_1^2\beta_2}
\frac{d\beta_1\wedge d\beta_2}{\beta_1\beta_2}.
\end{multline}
Evaluating corresponding residues we get
\begin{eqnarray}\label{PropEikonalBtermNMHV}
\hat{A}_{34}[B]&=&\frac{1}{\hat{\kappa}_{2,L}^*}
\frac{\langle5p_2\rangle^4}{\langle\hat{6}5\rangle\langle5p_2\rangle\langle p_2\hat{6}\rangle}
\times\frac{1}{\langle p_2|k_1+k_2|p_2]}\times\frac{1}{\hat{\kappa}_{2,R}^*}
\frac{\langle1p_2\rangle^4}{\langle12\rangle\langle2p_2\rangle\langle p_21\rangle},
\end{eqnarray}
where
\begin{eqnarray}\label{KappaShiftedNMHV6}
\hat{\kappa}_{2,L}^*=\frac{\langle p_2|q_5+\hat{q}_6|\xi]}{[p_2\xi]},~
\hat{\kappa}_{2,R}^*=\frac{\langle p_2|q_2+\hat{q}_1|\xi]}{[p_2\xi]}.
\end{eqnarray}
The action of $\hat{A}_{12}\circ \hat{A}_{56}$ can be evaluated in similar fashion and finally we arrive at
\begin{eqnarray}
\hat{A}_{12}\circ \hat{A}_{56}\circ\hat{A}_{34}[B]=A^*_{0+2}(g_2^*, \hat{g}_3^*)\frac{1}{\langle p_2|k_1|p_2]}A^*_{0+2}(\hat{g}_1^*,g_2^*),
\end{eqnarray}
which is exactly $\mathbb{B}$ term in the off-shell BCFW recursion \cite{vanHamerenBCFW1} for $A^*_{0+3}(g_1^*,g_2^*,g_3^*)$  amplitude.

So we see the pattern here: the contributions $\mathbb{C}$ and $\mathbb{D}$ from the off-shell BCFW recursion for $A^*_{0+3}(g_1^*,g_2^*,g_3^*)$ amplitude are reproduced from the on-shell BCFW recursion for $A_{3,6}(1^-,2^+,3^-,4^+,5^-,6^+)$ when gluing operator is acting on three point $\mbox{MHV}_3$ or $\overline{\mbox{MHV}}_3$ sub-amplitudes (with degenerate kinematics), while $\mathbb{B}$ contribution is
reproduced by the action of gluing operator on both sides of the BCFW bridge, see Fig. \ref{BCFW6point}.

These observations can be immediately generalized to the situation with arbitrary on-shell amplitude. One can obtain the off-shell
BCFW recursion for $A^*_{0+n}(g_1^*,\ldots,g_n^*)$ with the shift
of the off-shell momenta $k_1$ and $k_n$ from the BCFW recursion for $A_{n,2n}(1^-,2^+,\ldots,(2n)^{+})$ on-shell amplitude represented by $[1^-,(2n)^+\rangle$ shift. Indeed, the terms $\mathbb{C}$ and $\mathbb{D}$
are reproduced when gluing operator acts on $\mbox{MHV}_3$ or $\overline{\mbox{MHV}}_3$ on-shell amplitudes. Repeating the steps identical to the previous discussion (see (\ref{Dterm})) we get:
\begin{multline}
\hat{A}_{2n-1~2n}\circ \ldots\circ\hat{A}_{12}\left [A_{n-1,2n-1}\left(3^-, \ldots , \widehat{(2n)}^+, \hat{P}^+\right)\frac{1}{q_{1,2}^2}
A_{2,3}(\hat{1}^-,2^+,-\hat{P}^-)\right]\\
=\frac{1}{\kappa_1}A_{1+(n-1)}^*(g^*_2, \ldots , \hat{g}^*_n, \hat{P}^+),
\end{multline}
where
\begin{eqnarray}
|\hat{P}\rangle=\left(x(p_n)|p_1\rangle-
\frac{\kappa_1^*}{\langle p_n p_1\rangle}|p_n\rangle\right),\quad
|\hat{P}]=|p_1].
\end{eqnarray}
Similarly for $\overline{\mbox{MHV}}_3$ we have:
\begin{multline}
\hat{A}_{2n-1~2n}\circ \ldots\circ\hat{A}_{12}\left [A_{1,3}\left((2n-1)^-,\widehat{(2n)}^+,-\hat{P}^+\right)\frac{1}{q_{2n-1,2n}^2}
A_{n,2n-1}(\hat{1}^-,\ldots,\hat{P}^-)\right] \\
= \frac{1}{\kappa_n^*}A_{1+(n-1)}^*(\hat{g}^*_1,...,g^*_{n-1},\hat{P}^-),
\end{multline}
with
\begin{eqnarray}
|\hat{P}]=\left(x(p_1)|p_n]-
\frac{\kappa_1}{[p_n p_1]}|p_1]\right),\quad
|\hat{P}\rangle=|p_n\rangle.
\end{eqnarray}

When gluing operator $\hat{A}_{ii+1}$ acts on legs separated by the BCFW bridge the $\mathbb{B}$ type contribution is reproduced. In this case
the on-shell BCFW shift $z$ is replaced by
\begin{eqnarray}\label{zShiftBtermGeneral}
z=\frac{\langle p_i|k_1+\ldots+k_i|p_i]}{\langle p_1\, p_i\rangle[p_i\, p_{n}]},
\end{eqnarray}
and the factor $1/q_{1,i}^2$ is replaced by
\begin{eqnarray}
\frac{1}{\langle p_i|k_1+\ldots+k_i|p_i]}
\end{eqnarray}
times $\beta_i$ factors. The explicit proof that such contribution in general case gives us the $\mathbb{B}$ term can be done by
induction and can be sketched as follows: one can decompose each $A_{k_i,n_i}$ on-shell amplitude in individual BCFW terms via on-shell diagram representation \cite{AmplitudesPositiveGrassmannian} into combination of $\mbox{MHV}_3$
and $\overline{\mbox{MHV}}_3$ vertexes (on-shell diagrams). The action of the gluing operator on such on-shell diagrams
was considered in the previous section. After that one have to reassemble $A^{*}_{2,2+1}$, $\mbox{MHV}_3$
and $\overline{\mbox{MHV}}_3$ amplitudes together. As the result one can obtain that:
\begin{multline}
\ldots \circ\hat{A}_{ii+1}\circ \ldots \left [A_{k_1,n_1}((i+1)^+, \ldots , \widehat{(2n)}^+, \hat{P}^-)\frac{1}{q_{1,i}^2}
A_{k_2,n_2}(\hat{1}^-,\ldots,i^-,-\hat{P}^+)\right] \\
= A^*_{0+i}(\hat{g}_{1}^*, \ldots ,g_i^*)\frac{1}{\langle p_i|k_1+\ldots+k_i|p_i]}A^*_{0+(n-i)}(g_{i}^*, \ldots ,\hat{g}_n^*).
\end{multline}
Also, presented above relation is implicitly guaranteed by the Grassmannian integral representation of the on-shell and off-shell amplitudes. It was shown \cite{ambitwistorReggeons,ambitwistorFormfactors} that the latter could be easily related with each other by means of the same gluing operations (see (\ref{GluingGrassmannianIntegrals} and Appendix \ref{appA}) which imply
similar relation for the individual residues of the top-forms. We see that the gluing operator transforms ordinary
$1/P^2$ propagator type poles of the on-shell amplitudes into eikonal ones, when external legs, on which the gluing operator acts, are separated by the BCFW bridge (see Fig. \ref{generalBCFW} B ). The value of the BCFW shift parameter $z$ is adjusted accordingly to match the off-shell BCFW recursion term $\mathbb{B}$.

All other contributions (see Fig. \ref{generalBCFW}) reproduce $\mathbb{A}$ type terms from \cite{vanHamerenBCFW1},
which also can be shown by induction. Indeed it is easy to see that the pole factor
remains of the same $1/P^2$ type in this case, which corresponds to the propogators of the on-shell states of $\mathcal{N}=4$ SYM, and the value of $z$ is adjusted to match $\mathbb{A}$
term of the off-shell BCFW.

The fact that one can transform each individual term in the BCFW recursion for the on-shell amplitudes into
terms of the BCFW recursion for the Wilson line form factors using gluing operators $\hat{A}_{i,i+1}$ in fact is not (very)surprising and
in some sense trivial.
Indeed, as was mentioned before, if the relation (\ref{GluingGrassmannianIntegrals}) holds on the level of the Grassmannian integrals (top-cell diagrams) then
it likely will hold for the individual residues (boundaries of top-cells) as well\footnote{This can be explicitly seen for some particular case considering integration contours for tree level amplitudes $L^3_{n+2}$ and $\Omega^3_{n+2}$, and
probably can be easily generalised for the case of arbitrary number of Wilson line insertions and arbitrary value of $k$ \cite{ambitwistorFormfactors}. Here however we avoided considerations of integration contours completely.}. And it is also natural, that in the case of $\Omega_{n+2}^k$, and its generalisations to multiple Wilson line insertions, the residues of $\Omega_{n+2}^k$ can be identified with the individual BCFW terms of
recursion for the off-shell amplitudes in full analogy with the the on-shell case.

However observation that one can transform the BCFW recursion for the on-shell amplitudes into
the BCFW recursion for the Wilson line form factors at tree level opens up exiting possibility.
It is known that one can formulate the BCFW recursion not only for
the tree level on-shell amplitudes in $\mathcal{N}=4$ SYM theory but for the loop integrands as well
\cite{AllLoopIntegrandN4SYM,AmplitudesPositiveGrassmannian}.
\emph{If we can transform on-shell BCFW recursion into off-shell one at tree level, then what about recursion for the loop integrands} ?

The next sections will be dedicated to discussion of this question. Namely we will investigate
what will happen if we apply the analog of the gluing operators to the BCFW recursion
for $\mathcal{N}=4$ SYM loop integrands of the on-shell amplitudes. Our ultimate goal is to present arguments that using the gluing operator one can transform the integrands of
the on-shell amplitudes into the integrands of the Wilson line form factors at any given number of loops and external states.

\section{Gluing operation in momentum twistor space}\label{GluingMomentumTwistorsSection}

The integrands of the on-shell amplitudes in the planar limit are naturally formulated using momentum super twistors variables, in particular this is the case for theis BCFW recursion representation \cite{AllLoopIntegrandN4SYM}. So, to proceed with our main goal we need to fprmulate our gluing operation in momentum twistor variables as well.

\begin{figure}[t]
	\begin{center}
		\epsfxsize=6cm
		\epsffile{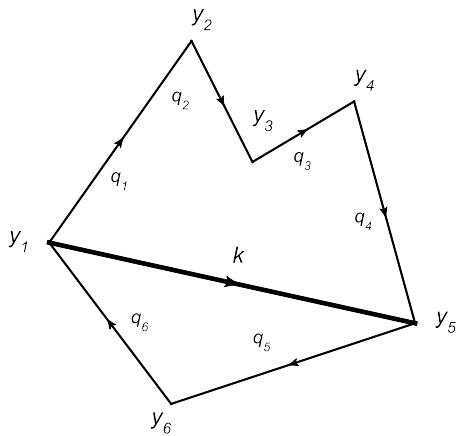}
	\end{center}\vspace{-0.2cm}
	\caption{Momenta and dual coordinates in the case of the amplitude with one off-shell and $n=4$ on-shell legs. In contrast to the case of the on-shell amplitudes, the $n$ on-shell momenta do not add up to zero but to the off-shell gluon momentum $k$: $q_1+\ldots+q_4=k$, which in its turn can be decomposed as a pair of auxiliary on-shell momenta $k=q_5+q_6$.}\label{fig: dual contour}
\end{figure}

To do this, let us recall how the momentum twistor variables are introduced. We start with so called zone super variables or dual super coordinates $y_i$ and $\vartheta_i$ which are related with on-shell momenta and its supersymmetric counterpart as \cite{Henrietta_Amplitudes,MomentumTwistors}:
\begin{eqnarray}
q_i=\vll_i \vlt_i = y_{i+1} - y_i, \quad \vll_i \vle_i = \vartheta_{i+1} - \vartheta_i.
\end{eqnarray}
The introduction of dual super coordinates helps to trivialize the conservation of super momentum \cite{Henrietta_Amplitudes,MomentumTwistors} and Fig. \ref{fig: dual contour} shows the momentum conservation geometrically for the case of $n=4$ on-shell and one off-shell momenta as an example. There we have a contour in the dual space formed by on-shell particles momenta together with two auxiliary on-shell momenta  $5$ and $6$ used to describe off-shell momentum.

The momentum super twistor variables $\mathcal{Z}_i = (\vll_i ,\mu_i , \vle_i)$ \cite{MomentumTwistors} are then defined through the following incidence relations
\begin{eqnarray}
\mu_i = \vll_i y_i = \vll_i y_{i+1},\quad \vlet_i = \vll_i\vartheta_i = \vll_i\vartheta_{i+1} .
\end{eqnarray}
The bosonic part of $\mathcal{Z}_i$ will be labeled as $Z_i=(\vll_i ,\mu_i)$.
Inverting presented above relations we get
\begin{equation}
\vlt = \mu\cdot Q\, ,\quad \vlet = \vle\cdot Q\, ,\quad  Q_{ij} = \frac{\delta_{i-1\, j}\abr{i\, i+1} + \delta_{i j}\abr{i+1\, i-1} + \delta_{i+1\, j}\abr{i-1\, i}}{\abr{i-1\, i}\abr{i\, i+1}}. \label{fromMomentumTwistors}
\end{equation}
Here $\vlt\equiv (\vlt_1 \cdots \vlt_n)$,  $\vlet\equiv (\vlet_1 \cdots \vlet_n)$ and it is assumed that $\sum_{i=1}^n q_i = 0$.
The transition from momentum twistors to helicity spinors could be performed with a formula like\footnote{The matrix $\tilde{Q}_{i j}$ is a formal inverse of singular map $Q_{i j}$, see \cite{LocalPhysicsGrassmanian,BCFWmathematica} for details.}:
\begin{equation}
\mu = \tilde{Q}\cdot \vlt\, ,\quad \vle = \tilde{Q}\cdot \vlet\, ,\quad \tilde{Q}_{i j} = \begin{cases}
\abr{ j\, i}\;\; \mbox{if}\; 1 < j < i\\
0\quad\;\;\; \mbox{otherwise}
\end{cases} \label{toMomentumTwistors}
\end{equation}
Note, that momentum super twistors trivialize both on-shell condition $q_i^2 = 0$ and mentioned above conservation of super momentum.

To construct the gluing operator acting in momentum twistor space let us recall that the initial gluing operator in helicity spinor variables  can be represented as an action of two consecutive BCFW bridges times some regulator\footnote{We call inverse soft factor (\ref{SoftFactor}) regulator because it makes soft holomorphic limit with respect to one of the auxiliary on-shell momenta, which encodes off-shell one, regular \cite{vanHamerenBCFW1,offshell-1leg}.} factor (\ref{GluibgOperAsBCFWBridges}).
The BCFW bridge operators can also be defined in momentum twistor space using special version of the on-shell diagrams \cite{AmplituhedronFromMomentumTwistors}. The action of $[i,i+1\rangle$ the BCFW shift bridge operator $br(\hat{i},i+1)$ in momentum twistor representation on the function $Y$ of $\{\ZZ_i\}_{i=1}^n$ variables is given by \cite{AllLoopIntegrandN4SYM,AmplituhedronFromMomentumTwistors}:
\begin{equation}
Y' (\ZZ_1,\ldots , \ZZ_n) = br(\hat{i},i+1)\left[ Y (\ZZ_1,\ldots , \ZZ_n)\right]\equiv \int\frac{d c}{c} Y (\ZZ_1,\ldots , \hat{\ZZ}_i, \ldots \ZZ_n),
\end{equation}
where $Y$, $Y'$ are both functions of $n$ momentum super twistors variables and $\hat{\mathcal{Z}}_i = \mathcal{Z}_i + c \mathcal{Z}_{i+1}$. We also do not require $Y$
and $Y'$ to be Yangian invariants.

As new on-shell diagrams in momentum
twistor space are no longer built from ordinary $\mbox{MHV}_3$ and $\overline{\mbox{MHV}}_3$ vertexes (amplitudes), then in the definition of the gluing operator we will have, in principle, to change the form of regulator factor. So, to construct the gluing operator in momentum twistors we will consider the following ansatz:
\begin{eqnarray}
\hat{A}_{i-1,i}^{m.twistor}\left[\ldots\right]=N~br(\hat{i},i+1)\circ br(\widehat{i+1},i) \left[M~\ldots\right]
\end{eqnarray}
with two unknown rational functions of helicity spinors $\lambda_i$ (first components of momentum twistors) - measure $M$ and normalization coefficient $N$.  To fix $N$, $M$ functions we require that
\begin{eqnarray}
\hat{A}_{n+1,n+2}^{m.twistor}\left[\mathcal{L}_{n+2}^k\right]=\omega^k_{n+2}\, ,
\end{eqnarray}
where ($k$ is N$^{(k-2)}$MHV degree and the use of an appropriate integration contour is assumed):
\begin{eqnarray}
\omega^k_{n+2} &=& \int
\frac{d^{(k-2)\times (n+2)}D}{\text{Vol}[GL(k-2)]}~Reg.~
\frac{\delta^{ 4 (k-2) | 4 (k-2)} (D\cdot \mathcal{Z})}{(1 \; \ldots \; k-2) \;\ldots \; (n+2 \; \ldots \; k-3)}\, , \nonumber\\
Reg.&=&\frac{1}{1+\frac{\la p_{n+1}\, \xi_{n+1}\ra}{\la p_{n+1}\, 1\ra}\frac{(n+2 \; 2 \; \ldots \; k-2)}{(1 \; \ldots \; k-2)}},
\label{GrassmannianMomentumTwistors}
\end{eqnarray}
is the momentum twistor Grassmannian integral representation for the ratio of amplitudes with one Willson line operator insertion $A^{*}_{k,n+1}/A^{*}_{2,n+1}$ and $\mathcal{L}_{n+2}^k$ is the Grassmannian representation for $A_{k,n+2}/A_{k=2,n+2}$ on-shell amplitude ratio:
\begin{eqnarray}
\mathcal{L}_{n+2}^k &=& \int
\frac{d^{(k-2)\times (n+2)}D}{\text{Vol}[GL(k-2)]}
\frac{\delta^{ 4 (k-2) | 4 (k-2)} (D\cdot \mathcal{Z})}{(1 \; \ldots \; k-2) \;\ldots \; (n+2 \; \ldots \; k-3)}.
\end{eqnarray}
That is our gluing operation should transform the Grassmannian integral representation of on-the shell amplitudes into corresponding Grassmannian integral representation for the off-shell amplitudes.  From this requirement we get
\begin{equation}
M=N^{-1}=S(i+1,i,i-1)\ ,
\end{equation}
where $S$ is the usual soft factor
\begin{eqnarray}
S(i+1,i,i-1)=\frac{\kappa_{i-1}^*\langle i-1\, i+1\rangle}{\langle i\, i+1\rangle\langle i-1\, i\rangle}\, .
\end{eqnarray}
Computation details can be found in Appendix \ref{appA}.

So, finally, we have the following expression for the gluing operation in momentum twistor space
\begin{equation}\label{GluingOperInMomentumTwistors}
\hat{A}_{i-1,i}^{m.twistor}\left[\ldots\right]=S(i+1,i,i-1)^{-1}~br(\hat{i},i+1)\circ br(\widehat{i+1},i) \left[S(i+1,i,i-1)\ldots\right].
\end{equation}
It may be, at first glance, surprising that here in momentum twistor space we used $[i,i+1\rangle$ BCFW shift and not $[i-1,i\rangle$ as for the gluing operation in the helicity spinors representation. In fact, mentioned before the two BCFW shifts are equivalent, see, for example, discussion in \cite{Henrietta_Amplitudes}. It is also assumed that in the construction of dual variables (dual contour)
we decompose any off-shell momentum $k_i$ that we encounter in a pair of (complex) on-shell
momenta as \cite{FormFactorsGrassmanians,offshell-1leg}:
\begin{equation}\label{Off-shellDecompForTwist}
	k_i=|p_i\ra\frac{\la \xi_i|k_i}{\la p_i\xi_i \ra}+|\xi_i\ra\frac{\la p_i|k_i}{\la p_i\xi_i \ra}.
\end{equation}
So we will have a pair of axillary momentum twistor variables $Z_i$ and $Z_{i+1}$ which encode
information about off-shell momenta $k_i$ (see Fig. \ref{fig: dual contour} as example). The same is also true for supersymmetric counterparts of $k_i$ momenta \cite{offshell-1leg}.

\begin{figure}[t]
	\begin{center}
		\epsfxsize=5cm
		\epsffile{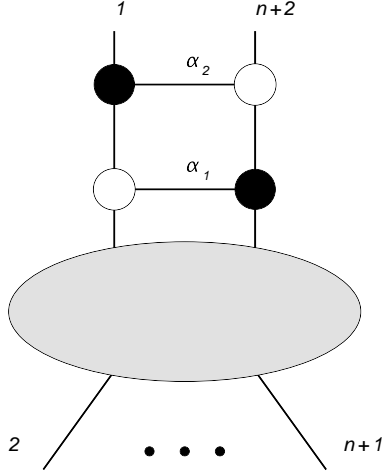}
	\end{center}\vspace{-0.2cm}
	\caption{Double BCFW bridge $br(1,\widehat{n+2})\circ br(n+2,\widehat{1})$.}\label{fig:AP(n+2)}
\end{figure}

Now, when we have an explicit definition of the gluing operator $\hat{A}_{ii+1}$
in momentum twistor space, let us proceed with particular applications of it. Hereafter we will drop $m.twistor$ subscript
to simplify the notations and hope that it will not lead to any confusion. First, consider the ratio
\begin{equation}
\mathcal{P}^{4(k-2)}_{n+2}(\mathcal{Z}_1
,...,\mathcal{Z}_{n+2})=A_{k,n+2}/A_{2,n+2}.
\end{equation}
Applying to it the gluing operation $\hat{A}_{i-1, i}$ we will have (see Fig. \ref{fig:AP(n+2)}):
\begin{multline}
\hat{A}_{i-1, i}\left[
\mathcal{P}^{4(k-2)}_{n+2}
\right] =  S(i+1, i, i-1)^{-1} \int \frac{d\alpha_1}{\alpha_1}
\frac{d\alpha_2}{\alpha_2}  \frac{\la i-1\, i+1 \ra + \alpha_1 \la i-1\, i\ra + \alpha_1\alpha_2 \la i-1\, i+1\ra}{\la i\, i+1\ra (\la i-1\, i\ra + \alpha_2 \la i-1\, i+1\ra)} \\
\times  \mathcal{P}^{4(k-2)}_{n+2} (\ldots, \mathcal{Z}_i + \alpha_2\mathcal{Z}_{i+1} , \mathcal{Z}_{i+1} + \alpha_1 \mathcal{Z}_i + \alpha_1 \alpha_2 \mathcal{Z}_{i+1}\, \ldots).
\end{multline}
Taking the residues at $\alpha_1 = 0$, $\alpha_2 = - \frac{\la i-1\, i\ra}{\la i-1\, i+1\ra}$ we finally get
\begin{equation}\label{ResultOfGluingMTwist}
\hat{A}_{i-1, i}\left[
\mathcal{P}^{4(k-2)}_{n+2}
\right] =  \mathcal{P}^{4(k-2)}_{n+2} (\ldots , \mathcal{Z}_i - \frac{\la i-1\, i\ra}{\la i-1\, i+1\ra}\mathcal{Z}_{i+1}, \mathcal{Z}_{i+1}, \ldots)\, ,
\end{equation}
which should be proportional to $A^{*}_{k,n+1}$ i.e.
\begin{equation}\label{ResultOfGluingMTwistAmpl}
\frac{A^{*}_{k,n+1}}{A^{*}_{2,n+1}}=\hat{A}_{i-1, i}\left[
\mathcal{P}^{4(k-2)}_{n+2}
\right],
\end{equation}
where the Wilson line operator insertion is positioned between the on-shell states with
numbers $i-2$ and $i+1$. If one has to consider several Wilson line operator insertions then one should apply the gluing operation several times similar to the examples considered in section \ref{BCFWfromGluing}.
\noindent
We see that application of the gluing operation in momentum twistor space
significantly simplifies compared to the helicity spinor case (\ref{Aoperator}) and amounts to just a shift of $i$-th momentum super twistor:
\begin{eqnarray}
	\mathcal{Z}_i^*=\mathcal{Z}_i - \frac{\la i-1\, i\ra}{\la i-1\, i+1\ra}\mathcal{Z}_{i+1},
\end{eqnarray}
which is similar to the BCFW shift. This shift, as expected, transforms ordinary $1/P^2_{k,j}$ propagators, which in momentum twistor space
are proportional to $1/\langle k-1 k j-1 j\rangle$, into eikonal ones (up to $\la i j \ra$ factors) if $i$
belongs to $(k-1,~k,~j-1,~j)$ set. As an example, let's consider the NMHV 6-point amplitude considered in the previous section. The term B in Fig. \ref{BCFW6point} before gluing
contained propagator pole which in dual variables is given by $1/x^2_{14}$ with $x^2_{14} \sim \la 61 34 \ra$. So
under the action of $\hat{A}_{34}$ we get $\la 61 34 \ra\mapsto \la 61 34^* \ra$, which in its turn can be transformed as:
\begin{eqnarray}
	\la 61 34^* \ra=\la 61 34 \ra +\frac{\la34 \ra}{\la35 \ra}\la 61 35 \ra
	=\frac{\la34 \ra}{\la35 \ra}\la3|(q_1+q_2+q_3)q_4|5\ra\sim\la p_2|q_1+q_2|p_2],
\end{eqnarray}
where we used explicit expressions for $q_3$ and $q_4$ ($k = q_3 + q_4$):
\begin{eqnarray}
	q_3=|p_2\ra\frac{\la \xi|k}{\la p_2\xi \ra},~ q_4=|\xi\ra\frac{\la p_2|k}{\la p_2\xi \ra},~k|p_2\ra=\kappa^*|p_2].
\end{eqnarray}
Comparing this expression to (\ref{PropEikonalBtermNMHV}) we see that $1/\la 61 34^* \ra$ is given exactly by the eikonal propagator. This observation can be easily generalized to the arbitrary $k,n$ case. So the whole analysis of section \ref{BCFWfromGluing} can be performed in the momentum twistor space with the identical result, which is accumulated in (\ref{ResultOfGluingMTwistAmpl}), (\ref{ResultOfGluingMTwist}) relations. We will not repeat it
here and will restrain ourselves to the consideration of some particular examples.

Namely, let's reproduce the results for $A^*_{3,4+1}/A^*_{2,4+1}$ and $A^*_{3,2+2}/A^*_{2,2+2}$ off-shell amplitudes using
the gluing operator. To do this we start with the ratio $\mathcal{P}^{4}_{6} = A_{3,6}/A_{2,6}$ of the on-shell amplitudes
\begin{equation}
\mathcal{P}^{4}_{6} = [12345] + [13456] + [12356]\, ,
\end{equation}
where as usual five-bracket is given by \cite{Henrietta_Amplitudes}:
\begin{eqnarray}
[i \; j \; k \; l \; m] = \frac{\delta^4 (\la i \; j \; k \; l\ra\vle_m + \text{cyclic permutation})}{\la i \; j \; k \; l\ra\la j \; k \; l \; m\ra\la k \; l \; m \; i\ra\la l \; m \; i \; j\ra\la m \; i \; j \; k\ra}
\end{eqnarray}
with four-brackets defined as
\begin{eqnarray}
\la i \; j \; k \; l\ra = \varepsilon_{A B C D} Z_i^A Z_j^B Z_k^C Z_l^D .
\end{eqnarray}
Then we have\footnote{It is assumed that the momentum super twistors $\mathcal{Z}_5$ and $\mathcal{Z}_6$ are sent to corresponding off-shell kinematics related to off-shell momenta of $g_5^*$ reggeized gluon.}
\begin{eqnarray}
\hat{A}_{5, 6}\left[
\mathcal{P}^{4}_{6}
\right] &=& \frac{1}{1+\frac{\la p_5 \xi_5\ra}{\la p_5 1\ra}\frac{\langle 1345\rangle}{\langle 3456\rangle}}[13456]+\frac{1}{1+\frac{\la p_5 \xi_5\ra}{\la p_5 1\ra}\frac{\langle 1235\rangle}{\langle 2356\rangle}}[12356]+[12345],\nonumber\\
\hat{A}_{5, 6}\left[
\mathcal{P}^{4}_{6}
\right]&=&\frac{A^*_{3,4+1}}{A^*_{2,4+1}}\left(\Omega_1,\ldots,\Omega_4,g^*_5\right),
\end{eqnarray}
and\footnote{We again assume corresponding off-shell kinematics for momentum super twistors $\mathcal{Z}_3$ - $\mathcal{Z}_6$ describing reggeized gluons $g_3^*$ and $g_4^*$.}
\begin{eqnarray}
\hat{A}_{3, 4}\circ \hat{A}_{5, 6}\left[
\mathcal{P}^{4}_{6}
\right] &=&
c_{35} [12345] + c_{36} [12356] + c_{46} [13456] ,  \nonumber \\
\hat{A}_{3, 4}\circ \hat{A}_{5, 6}\left[
\mathcal{P}^{4}_{6}
\right]&=&\frac{A^*_{3,2+2}}{A^*_{2,2+2}}\left(\Omega_1,\Omega_2,g^*_3,g^*_4\right),
\end{eqnarray}
with
\begin{eqnarray}
c_{35} = \frac{1}{1+\frac{\la p_3\xi_3\ra\la 1 2 3 5\ra}{\la p_3 p_4\ra\la 1234\ra}}, \quad
c_{36} = \frac{1}{1+\frac{\la p_4\xi_4\ra}{\la p_4 1\ra}\frac{\la 1 2 3 5\ra}{\la 2 3 5 6\ra}}, \quad
c_{46} = \frac{1}{1+\frac{\la p_3\xi_3\ra}{\la p_3 p_4\ra}\frac{\la 1 3 5 6\ra}{\la 1 3 4 6\ra}}
\frac{1}{1+\frac{\la p_4\xi_4\ra}{\la p_4 1\ra}\frac{\la 1 3 4 5\ra}{\la 3 4 5 6\ra}} . \nonumber \\
\end{eqnarray}
These results are in complete agreement with previously obtained results from the off-shell BCFW \cite{vanHamerenBCFW1} and Grassmannian integral representation \cite{offshell-1leg,offshell-multiple}.

In general the on-shell ratio function $\mathcal{P}_{k,n+2}=\mathcal{P}^{4(k-2)}_{n+2}(\mathcal{Z}_1
,...,\mathcal{Z}_{n+2})$ can be found for fixed $n$ and $k$ via the solution of the on-shell BCFW recursion in momentum twistor space \cite{AllLoopIntegrandN4SYM}:
\begin{multline}\label{BCFWTwistorTrees}
\mathcal{P}_{k,n}(\mathcal{Z}_1
,...,\mathcal{Z}_{n}) =  \mathcal{P}_{k,n-1}(\ZZ_1 , \ldots , \ZZ_{n-1}) \\
+ \sum_{j=2}^{n-2} [j-1, j, n-1, n, 1] \mathcal{P}_{k_1,n+2-j} (\ZZ_{I_j}, \ZZ_j, \ZZ_{j+1}, \ldots , \hat{\ZZ}_{n_j}) \mathcal{P}_{k_2, j} (\ZZ_{I_j}, \ZZ_1, \ZZ_2, \ldots , \ZZ_{j-1})\, ,
\end{multline}
where\footnote{$(i, j)\cap (k, p, m) \equiv \ZZ_i\la jkpm \ra+\ZZ_j\la ikpm \ra  $} $\ZZ_{n_j}= (n-1, n)\cap (1, j-1, j)$, $\hat{\ZZ}_{I_j}= (j-1, j)\cap (1, n-1, n)$,
$k_1+k_2+1=k$. We will make more comments about the structure of this recursion relation in the next chapter. From practical point of view the easiest way to compute Wilson line form factor with $f$ on-shell states and $m$ Wilson line operator insertions is to solve (\ref{BCFWTwistorTrees})  for $n=f+2m$ and then apply $m$ gluing operators via (\ref{ResultOfGluingMTwist}) rule.

Now, when we have the definition of the gluing operator $\hat{A}_{ii+1}$ in
momentum  twistor space and some practice with the tree level answers we are ready to consider loop integrands.

\section{Loop integrands}\label{IntegrandsSection}

The natural way to define planar loop integrands unambiguously is to use momentum twistors or dual coordinates. The loop integrand $I_{k,n}^L$  for on-shell $L$-loop amplitude $A_{k,n}^{L}$ in this language is defined as\footnote{Here by dividing on MHV amplitude we mean
that we are factoring out $\la 12\ra \ldots \la n1 \ra$ product and dropping momentum
conservation delta function.}
\begin{equation}
A_{k,n}^{(L)}/A_{2,n}^{(0)} = \int_{reg}\prod_{m=1}^L d^4 l_m I_{k,n}^{(L)} (\ZZ_1,\ldots , \ZZ_n ; {l_1, \ldots ,l_L})\, , \label{integranddef}
\end{equation}
where momentum super twistors $\ZZ_1, \ldots , \ZZ_n$ describe kinematics of external particles and $reg$ stands for regularization needed by loop integrals. Here $I_{k,n}$ is a rational function of both loop integration and external kinematical variables. Moreover, $I_{k,n}$ is cyclic in external momentum super twistors. It is also assumed that loop integrand is completely symmetrized in loop variables $l_1,\ldots , l_L$. Rewriting the latter in terms of bi-twistors ( $l_m\equiv (A_m B_m)\equiv (A B)_m$) the loop integration measure takes the form \cite{AllLoopIntegrandN4SYM}:
\begin{equation}
d^4 l = \la A B d^2 A\ra \la A B d^2 B\ra = \frac{d^4 Z_A d^4 Z_B}{\text{Vol}[GL(2)]}\, ,
\end{equation}
where we dropped out factors $\abr{\lambda_A\, \lambda_B} = \la\ZZ_A\ZZ_B I_{\infty}\ra$ as the integrands in $\mathcal{N}=4$ SYM are always dual conformal invariant. Here $I_{\infty}$ denotes infinity bi-twistor \cite{MomentumTwistors}. The integral over the line $(A B)$ is given by the integrals over the points $Z_A$, $Z_B$ modulo $GL(2)$ transformations leaving them on the same line.

\subsection{BCFW for integrands of  Wilson lines form factors and correlation functions}
\begin{figure}[t]
	\begin{center}
		\epsfxsize=12cm
		\epsffile{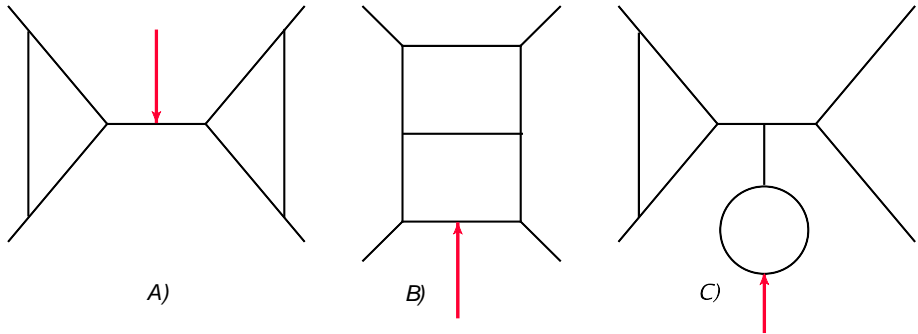}
	\end{center}\vspace{-0.2cm}
	\caption{Different possible types of pole contributions to loop BCFW recursion. Here we depicted scalar integrals for two loop $n=4$ example. Red arrows indicate propagators which we are cutting when evaluating residues. Term C) is actually absent in $\mathcal{N}=4$ SYM case as well as A).}\label{LoopIntPoles}
\end{figure}
Now let us see what modifications occur to the on-shell integrand BCFW recursion in the off-shell case. The loop-level BCFW for on-shell amplitudes in $\mathcal{N}=4$ SYM was worked out in detail in \cite{AllLoopIntegrandN4SYM} (see also \cite{Boels:2010nw,Boels:2016jmi} for situation with less SUSY) and the result for $\hat{\ZZ}_n = \ZZ_n + w \ZZ_{n-1}$ shift reads
\begin{multline}\label{BCFWInLoops}
I_{k,n}^{(L)} =  I_{k,n-1}^{(L)}(\ZZ_1 , \ldots , \ZZ_{n-1}) \\
+ \sum_{j=2}^{n-2} [j-1, j, n-1, n, 1] I_{k_1,n+2-j}^{(L_1)} (\ZZ_{I_j}, \ZZ_j, \ZZ_{j+1}, \ldots , \hat{\ZZ}_{n_j}) I_{k_2, j}^{(L_2)} (\ZZ_{I_j}, \ZZ_1, \ZZ_2, \ldots , \ZZ_{j-1}) \\
+ \int \frac{d^{4|4}\ZZ_A d^{4|4}\ZZ_B}{\text{Vol}[GL(2)]}\int_{GL(2)} [A,B,n-1,n,1] I_{k+1,n+2}^{(L-1)} (\ZZ_1 , \ZZ_2 , \ldots , \hat{\ZZ}_{n_{AB}}, \ZZ_A, \ZZ_B)\, ,
\end{multline}
where $\hat{\ZZ}_{n_j}= (n-1, n)\cap (1, j-1, j)$, $\ZZ_{I_j}= (j-1, j)\cap (1, n-1, n)$, $\hat{\ZZ}_{n_{AB}}= (n-1, n)\cap (A, B, 1)$ and $k_1+k_2+1=k$.
The $\int_{GL(2)}$ integral is defined as follows. First we set $\ZZ_A\to \ZZ_A + \alpha \ZZ_B\equiv \ZZ'_A$ and $\ZZ_B\to \ZZ_B + \beta \ZZ_A\equiv \ZZ'_B$, which is equivalent to moving points $\ZZ_A$ and $\ZZ_B$ without changing the line they span. Then we calculate composite residue in $\alpha$, $\beta$ such that $\la A', 1, n-1, n\ra\to 0$ and $\la B', 1, n-1, n\ra\to 0$, what is equivalent to taking points $A'$, $B'$ to lie on the plane $\la 1, n-1, n \ra$:
\begin{equation}
\int_{GL(2)}\equiv \int_{\la A', 1, n-1, n\ra\to 0}d\alpha
\int_{\la B', 1, n-1, n\ra\to 0} d\beta\, (1-\alpha\beta)^2
\end{equation}
Taking the residue as above is equivalent to setting $\ZZ'_A, \ZZ'_B$ to $(A, B)\cap (1, n-1, n)$ and the Jacobian factor $(1-\alpha\beta)^2$ makes poles in $\alpha , \beta$ simple.

\begin{figure}[t]
	\begin{center}
		\epsfxsize=8cm
		\epsffile{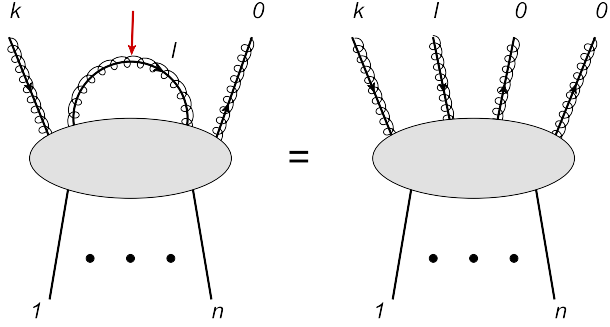}
	\end{center}\vspace{-0.2cm}
	\caption{Evaluation of residue in the pole (cut) of eikonla loop propagator.
	Red arrow indicates which propagator we are cutting.}\label{ForwardLimitWilsonLineFF}
\end{figure}

Next, let us make some comments about the origin of different terms in (\ref{BCFWInLoops}). The first two terms, namely
\begin{multline}\label{BCFWInLoops1part}
I_{k,n-1}^{(L)}(\ZZ_1 , \ldots , \ZZ_{n-1})
+ \sum_{j=2}^{n-2} [j-1, j, n-1, n, 1] I_{k_1,n+2-j}^{(L_1)} (\ZZ_{I_j},  \ldots , \hat{\ZZ}_{n_j}) I_{k_2, j}^{(L_2)} (\ZZ_{I_j},  \ldots , \ZZ_{j-1})
\end{multline}
originate from the poles in the BCFW shift parameter $w$ coming from
propagators which does not contain loop momentum dependence:
\begin{eqnarray}
	\la i-1 i n-1\hat{n}(w) \ra=0.
\end{eqnarray}
that is from propagators connecting loop integrals, see Fig. \ref{LoopIntPoles} A.
These contributions are identical both at tree and loop level. The term containing $GL(2)$ integration
\begin{equation}\label{BCFWInLoops2part}
\int_{GL(2)} [A,B,n-1,n,1] I_{k+1,n+2}^{(L-1)} (\ZZ_1 , \ZZ_2 , \ldots , \hat{\ZZ}_{n_{AB}}, \ZZ_A, \ZZ_B)
\end{equation}
is present only at the loop level. It originates from the poles in the BCFW shift parameter $w$ coming from propagators containing loop momenta \cite{AllLoopIntegrandN4SYM}, see Fig. \ref{LoopIntPoles} B.
\begin{figure}[t]
	\begin{center}
		\epsfxsize=8cm
		\epsffile{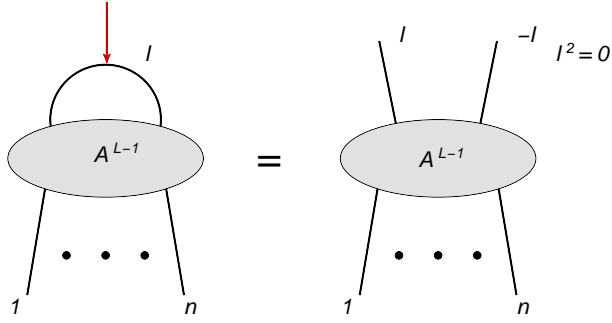}
	\end{center}\vspace{-0.2cm}
	\caption{Evaluation of residue at the pole of loop propagator (cut) resulting in the forward limit. Red arrow indicates which propagator we are cutting.}\label{ForwardLimitAmplitudes}
\end{figure}
At $L$ loop level for $n$ point amplitude the residue at such pole corresponds to the so called forward limit of $L-1$ loop
$n+2$ point amplitude. Indeed, if we consider $L$ loop integrand\footnote{Here we assume some specific ``appropriate'' choice of loop momenta. The corresponding ambiguity in the choice of loop momenta can be removed \cite{AllLoopIntegrandN4SYM} if one considers dual (or momentum twistor) variables and planar limit, which we are interested in.}
of some amplitude $I^{(L)}_n(\{p_1,\ldots,p_n\},l_1,\ldots,l_L)$, where $\{p_1,\ldots,p_n\}$ are external momenta and consider residue
at the pole $1/l^2_L$ corresponding to $L$'th loop integration
we will get (see Fig. \ref{ForwardLimitAmplitudes})
\begin{equation}\label{FL}
Res_{l_L^2=0}~I^{(L)}_n \sim I^{(L-1)}_{n+2}(\{p_1,\ldots,p_n,-l_L,l_L\},l_1,\ldots,l_{L-1}).
\end{equation}
In momentum twistor space residue can be evaluated as follows. For simplicity
let's consider $L=1$ example to make formulas more readable. The generalization for general $L$ is trivial.  The $n$-point amplitude integrand is the function of the following variables $I^{(1)}_n(\{\ZZ_1\ldots,\ZZ_n\},\ZZ_A,\ZZ_B)$. The residue
at the point (we consider  $\hat{\ZZ}_n = \ZZ_n + w \ZZ_{n-1}$ shift and take residue with respect to $w$ parameter)
 \begin{equation}
\la AB 1\hat{n}(w)\ra=0
\end{equation}
is given by:
\begin{equation}\label{FLTwist}
Res_{\la AB 1\hat{n}\ra=0}~I^{(1)}_n\sim A^{tree}_{n+2}(\ZZ_1,\ldots,\hat{\ZZ}_n,\hat{\ZZ}_B,\hat{\ZZ}_B),
\end{equation}
where
\begin{eqnarray}\label{FLTwist1}
\hat{\ZZ}_n&=&(n-1, n)\cap (A, B, 1),\nonumber\\
\hat{\ZZ}_B&=&(A, B)\cap (n-1, n, 1).
\end{eqnarray}
This is analog of (\ref{FL}) in
momentum twistor space, see also Fig. \ref{ForwardLimitWilsonLineFF}.
The first expression for $\hat{\ZZ}_n$ solves $\la AB 1\hat{n}\ra=0$. The second expression for $\hat{\ZZ}_B$ is the consequence of the first one and the forward limit. See \cite{Henrietta_Amplitudes} for detailed derivation and discussion. The expression (\ref{FLTwist}) in this limit could be obtained from the expression for  $A^{tree}_{n+2}(\ZZ_1,\ldots,\hat{\ZZ}_n,\ZZ_A,\hat{\ZZ}_B)$ at general kinematics\footnote{General in a sense that there are no collinear twistors in contrast to (\ref{FLTwist}).} by introducing $GL(2)$ integration with
$[A,B,n-1,n,1]$ weight (\ref{BCFWInLoops2part}).

Now let's see how similar to (\ref{BCFWInLoops}) the recurrence relation for Wilson line form factors can be constructed. Let's
consider integrand $I^{*(L)}_{k,n+1}$
of $A^{*(L)}_{k,n+1}(\Omega_1^*,\ldots,\Omega_n,g^*_{n+1})$ Wilson line form factor. As one will try to reconstruct it via $\hat{\ZZ}_i = \ZZ_i + w \ZZ_{i-1}$ shift he/she will encounter two types of contributions. The first type will be given by the residues with respect to propagators which does not contain loop momentum dependence. These can be considered along the same lines as in sections \ref{offshellBCFW} and \ref{GluingMomentumTwistorsSection}. The second type of contribution is the residues with respect to propagator poles with loop momentum dependence. Now in contrast to the case of on-shell amplitudes we have two types of propagator poles. Ordinary $1/l^2$ poles and eikonal ones $1/\la p| l| p]$. To simplify discussion let's consider one-loop case. Generalization to higher loops can be easily done by induction. The residue evaluation with respect to $1/l^2$ poles is identical to the on-shell amplitudes case and is given by forward limit of tree level Wilson line form factor with $n+2$ on-shell legs ($k$ as usual is off-shell momentum with direction $p$ and $\{q_1,\ldots,q_n\}$ are on-shell momenta):
\begin{equation}\label{FL_p2_pole_WL}
Res_{l_L^2=0}~I^{*(L=1)}_{n+1} \sim A^{*(tree)}_{(n+2)+1}(\{q_1,\ldots,q_n,-l_L,l_L\},\{p,k\}).
\end{equation}
The terms which include eikonal propagator pole residue are a little more complicated. Surprisingly, here similar to the on-shell case we also have forward like limit. For example, consider Wilson line form factor at one loop level $I^{*(1)}_n(\{q_1,\ldots,q_n\},\{k,p\},l)$. Here once again $\{q_1,\ldots,q_n\}$ are on-shell momenta and $k$ is off-shell momentum with direction $p$. Using decomposition (\ref{Off-shellDecompForTwist}) we can decompose our off-shell momentum into pair of on-shell momenta $k'=|p\ra k|\xi\ra /\la p \xi \ra$, $k''=|\xi\ra k|p\ra / \la  \xi p \ra$ and formally write this integrand as
$I^{*(1)}_n(\{q_1,\ldots,q_n,k'',k'\},l)$. Considering residue for the pole
$1/\la p|l|p]$ we enforce on loop momentum $l$ condition $\la p|l|p]=0$, $l^2\neq 0$ (see also (\ref{EikonalOnShellCondition}) and discussion there). This results in
(see Fig. \ref{ForwardLimitWilsonLineFF})
\begin{equation}\label{FLforWL}
Res_{\la p|l|p]=0}~I^{*(1)}_{n+1}\sim A^{*}_{n+2}(\{q_1,\ldots,q_n,k'',k',l',l''\}),
\end{equation}
where $l'=|p\ra l|\xi'\ra / \la  \xi' p \ra$, $l''=|\xi'\ra l|p\ra / \la  \xi' p \ra$. Now
using the freedom in the choice of $|\xi'\ra$ one can set $l'$ and $k'$ collinear to each other:
\begin{equation}
(l')^{\mu} = -(k')^{\mu}/ \la  \xi' p \ra
\end{equation}
up to scalar factor $\la  \xi' p \ra$. This resembles the on-shell forward limit kinematics of (\ref{FL}) for $n+4$ point off-shell amplitude. So presumably the residue with respect to Wilson line propagators can, in principle, be evaluated in momentum twistor space along the same lines as (\ref{FLTwist}) and (\ref{FLTwist1}).
\begin{figure}[t]
	\begin{center}
		\epsfxsize=13cm
		\epsffile{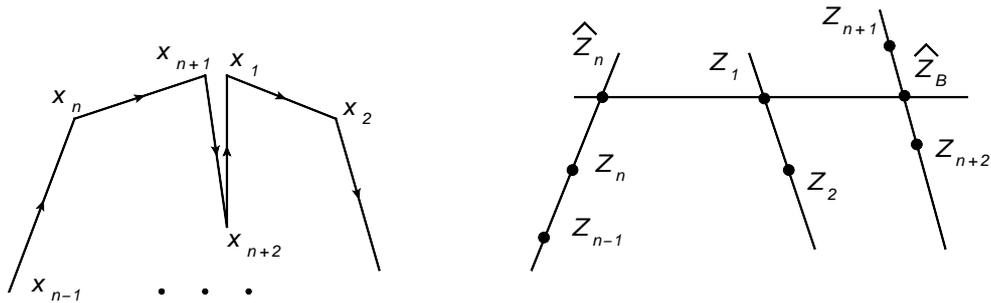}
	\end{center}\vspace{-0.2cm}
	\caption{Forward limit in momentum and momentum twistor space. In momentum space we are gluing $x_{n+1}$ with $x_{1}$ while keeping $x_{n}$ fixed in such
a way that $x_{1n}^2=0$. In momentum twistor space this equivalent to gluing
$Z_{n+1}$ and $Z_{n+2}$ with $\hat{Z}_B$. The same is also true for their supersymmetric counterparts.}\label{ForwardLimit}
\end{figure}
Consideration of this eikonal residue type, however, \emph{can be avoided entirely} if one will choose BCFW shift in such a way that $w$ parameter will not appear in eikonal propagators at all.

To see this let's consider once again one-loop case, that is we take the solution of (\ref{BCFWInLoops}) for $n$ external particles $I_{k,n}^{(L=1)}$ and apply gluing operator $\hat{A}_{n-1n}$ to it
\begin{eqnarray}\label{1loopI}
	I_{k,(n-2)+1}^{*(L=1)}=\hat{A}_{n-1n}[I_{k,n}^{(L=1)}].
\end{eqnarray}
We will assume that the tree level form factors and on-shell amplitudes are related as
\begin{eqnarray}\label{1loopA}
	\hat{A}_{n-1n}[A_{k,n}(\Omega_1\ldots,\Omega_n)]=A^*_{k,(n-2)+1}(\Omega_1\ldots,\Omega_{n-2},g^*_{n-1}).
\end{eqnarray}
What we are going to show now is that $I_{k,(n-2)+1}^{*(L=1)}$ will have appropriate factorization properties\footnote{That is the corresponding residue will be given by forward limit of tree level Wilson line form factor with $n+2$ on-shell states.} for one loop Wilson line form factor and that it can be obtained from recurrence relation similar to (\ref{BCFWInLoops}), where only poles of (\ref{FL_p2_pole_WL}) type will contribute. I.e. there always will be possibility to choose the BCFW shift in such a way that only $1/P^2$ type poles will contribute to recursion.

To show this let us consider all possible BCFW shifts in $I_{k,(n-2)+1}^{*(L=1)}$. But first let us note that in the case under consideration the pair of axillary momentum twistor variables $\ZZ_{n}$ and $\ZZ_{n-1}$ is used to encode information about off-shell momentum $k$ according to (\ref{Off-shellDecompForTwist}).
So, the only possible propagators which contain loop momentum and which will be affected by gluing operator $\hat{A}_{n-1n}$  are given by $\la AB n-1 n\ra$ and $\la AB n1 \ra$.
More accurately,  only $\la AB n-1 n\ra$ will be transformed into eikonal propagator $\la AB n-1 n^*\ra$ since $\la AB n^*1 \ra=\la AB n1 \ra$. Equivalently
one can note that due to the cyclical symmetry the only possible eikonal propagator with loop momentum dependence in
$A^{*(L=1)}_{k,(n-2)+1}$ will depend on $Z_A,Z_B,Z_{n-1},Z_{n},Z_1$ momentum twistors.

Now let's return to the shifts. If we shift $\ZZ_i$ as $\hat{\ZZ}_i = \ZZ_i + w \ZZ_{i-1}$ for $i=1,\ldots,n-2$, then the shift parameter $w$ will not affect the eikonal propagator and the corresponding residues with respect to $w$ can be evaluated according to (\ref{FL_p2_pole_WL}), so that the result will be given by the forward limit of the tree level Wilson line form factor with $n$ on-shell states. These is precisely the desired factorization property. For this form factor we also know that the relation (\ref{1loopA}) holds. So we see that in such cases the gluing operation indeed transforms solutions of (\ref{BCFWInLoops}) into Wilson line form factors similar to tree level.

As for the shifts involving $\ZZ_{n-1}$ and $\ZZ_{n}$,  we can always choose to shift $\hat{\ZZ}_n=\ZZ_n+w\ZZ_{n-1}$, so that the $w$ parameter drops out of $\la AB n-1 n^*\ra$ bracket and will remain only in $\la AB \hat{n}1 \ra$ bracket, which is again not affected by the action of $\hat{A}_{n-1n}$ gluing operator. This gives us
\begin{equation}\label{FLTwistWL}
Res_{\la AB \hat{n} 1\ra=0}~\hat{A}_{n-1n}[I^{(1)}_{n}]\sim\hat{A}_{n-1n}[A^{tree}_{n+2}]= A^{*}_{(n)+1}(\ZZ_1,\ldots,\ZZ_{n-1},\hat{\ZZ}_n^*,\hat{\ZZ}_B,\hat{\ZZ}_B),
\end{equation}
where
\begin{eqnarray}\label{FLTwistWL1}
\hat{\ZZ}_n^*&=&(n-1, n^*)\cap (A, B, 1),\nonumber\\
\hat{\ZZ}_B&=&(A, B)\cap (n-1, n, 1),\nonumber\\
\ZZ_n^*&=&\ZZ_n+\frac{\la p \xi \ra}{\la p 1\ra}\ZZ_1.
\end{eqnarray}
Here we see that $\hat{\ZZ}_n^*$ solves $\la AB \hat{n}^*1\ra=\la AB \hat{n}1\ra =0$ -- the same condition as in the case of the on-shell amplitudes. So once again we have appropriate factorization properties and we also see that the gluing operation indeed transforms solutions of
(\ref{BCFWInLoops}) into the Wilson line form factors.

Equivalently using the same arguments as above one can show that in $A^{*(L=1)}_{k,(n-2)+1}$ in pair $\ZZ_{n-1},\ZZ_{n}$
one can always choose to shift $\hat{\ZZ}_n=\ZZ_n+w\ZZ_{n-1}$ so that $w$ will
drop out from eikonal propagator. That is for all $\hat{\ZZ}_i$, which describe both on-shell and off-shell momenta, one can choose such shifts that will not affect eikonal propagators with loop momentum dependance and the corresponding recurrence relations will contain only contribution of (\ref{BCFWInLoops1part}) and (\ref{FL_p2_pole_WL}) type.

This considerations can be easily generalized by induction to arbitrary loop level and to arbitrary number of gluing operators applied. So we may conclude that application of $\hat{A}_{i-1i}$ to (\ref{BCFWInLoops})
will likely result in a valid recursion relation for loop integrands of Wilson line form factors (off-shell amplitudes) similar to tree level case. For example if we chose $i=n$, to match our previous considerations, we will get recurrence relation for the integrand $I_{k,(n-2)+1}^{*(L)}$ of Wilson line form factor when operator is inserted after on-shell state with number $n-2$:
\begin{multline}\label{BCFWInLoopsWLFF}
I_{k,(n-2)+1}^{*(L)} =  I_{k,n-1}^{(L)}(\ZZ_1 , \ldots , \ZZ_{n-1}) \\
+ \sum_{j=2}^{n-2} [j-1, j, n-1, n^*, 1] I_{k_1,n+2-j}^{(L_1)} (\ZZ_{I_j}, \ZZ_j, \ZZ_{j+1}, \ldots , \hat{\ZZ}_{n_j}^*) I_{k_2, j}^{(L_2)} (\ZZ_{I_j}, \ZZ_1, \ZZ_2, \ldots , \ZZ_{j-1}) \\
+ \int \frac{d^{4|4}\ZZ_A d^{4|4}\ZZ_B}{\text{Vol}[GL(2)]}\int_{GL(2)} [A,B,n-1,n^*,1] I_{k+1,n+2}^{(L-1)} (\ZZ_1 , \ZZ_2 , \ldots , \hat{\ZZ}_{n_{AB}}^*, \ZZ_A, \ZZ_B)\, ,
\end{multline}
where $\hat{\ZZ}_{n_j}= (n-1, n^*)\cap (1, j-1, j)$, $\ZZ_{I_j}= (j-1, j)\cap (1, n-1, n)$, $\hat{\ZZ}_{n_{AB}}= (n-1, n^*)\cap (A, B, 1)$ and $k_1+k_2+1=k$. $\ZZ_n^*$ is given by (\ref{FLTwistWL1}).
As before, to encode off-shell momenta we use twistor variables
with numbers $n-1$ and $n$. $p$ and $\xi$ are light-cone vectors entering $k_T$ - decomposition of this off-shell momentum $k$. Spinors $|p\ra$ and $|\xi\ra$ are obtained from corresponding vectors.

One can also skip the solution of this new recursion and apply $\hat{A}_{i-1i}$ directly to the solutions of on-shell recursion relation (\ref{BCFWInLoops}), that is to the on-shell integrands,
similar to the tree level case (\ref{ResultOfGluingMTwist}). In the next section we will consider such action using local form of integrands instead of non-local form produced directly by BCFW recursion.

At the end of this section we want to make the following note: in general forward limits may not be well defined \cite{AllLoopIntegrandN4SYM},
because \emph{on the level of integrands} one may encounter contributions from
tadpoles and bubble type integrals on external on-shell legs (see Fig. \ref{LoopIntPoles} C as an example). However, such contributions are absent
in $\mathcal{N}=4$ SYM on-shell amplitudes due to the enhanced SUSY cancellations \cite{AllLoopIntegrandN4SYM,Henrietta_Amplitudes}. Their analogs are also absent for the Wilson line form factors (off-shell reggeon amplitudes) - there are no tadpoles diagrams involving closed Wilson line propagators and bubbles on external Wilson line are also equal to 0 on integrand level (see Feynman rules in \cite{vanHamerenBCFW1}).

\subsection{Gluing operation and local integrands}

Now, following our discussion in the previous subsection we conclude that the integrands for the planar off-shell $L$-loop amplitudes could be obtained from the corresponding on-shell integrands by means of the same gluing procedure as was used by us at tree level. Namely, for reggeon amplitude with
$n$ reggeized gluons (Wilson line operator insertions) and no on-shell states $I_{k,0+n}^{* (L)}(g_1^*,\ldots,g_n^*)$ we should have:
\begin{multline}
I_{k,0+n}^{* (L)} = \hat{A}_{2n-1\, 2n}\circ\ldots \circ\hat{A}_{12}\Big[ I_{k, 2n}\Big] \\ = I_{k,2n}^{(L)}\left(
\ZZ_1 , \ZZ_2 - \frac{\abr{1\, 2}}{\abr{1\, 3}}\ZZ_3 \, \ldots , \ZZ_{2n-1} , \ZZ_{2 n} - \frac{\abr{2n-1\, 2n}}{\abr{2n-1 1}}\ZZ_1
\right). \label{integrand-gluing}
\end{multline}
Here it is assumed that $I_{k,0+n}^{* (L)}$ is normalized by $A_{2,0+n}^{* (0)}$ similar to the definition of on-shell integrands (\ref{integranddef}).
The loop integrands for reggeon amplitudes (Wilson line form factors) with on-shell states can be obtained from
(\ref{integrand-gluing}) by removing necessary number of $\hat{A}_{i-1i}$ operators.

The loop integrands produced by the BCFW recursion are non-local in general \cite{AllLoopIntegrandN4SYM}. However, it is still possible to rewrite the integrands in a manifestly local form\footnote{This procedure spoils the Yangian-invariance of each term in the on-shell case however.}. Moreover, one may choose as a basis the set of chiral integrals with unit leading singularities  \cite{AllLoopIntegrandN4SYM,LocalIntegrands}. The leading singularities are generally defined as the residues of a complex, multidimensional integrals of integrands in question over $\mathbb{C}^{4L}$, where $L$ is the loop order. The computation of residues for the integrands expressed in momentum twistors is then ultimately related to the classic Schubert problem in the enumerative geometry of $\mathbb{CP}^3$ \cite{LocalIntegrands}.  When the residues of integral associated to at least one of its Schubert problems are not the same then the integral is called chiral. In the case when the integral has at most one non-zero residue for the solutions to each Schubert problem then the integral is called completely chiral. If all non-vanishing residues are the same up to a sign then it is possible to normalize them, so that all residues are $\pm 1$ or $0$. The integrals with this property are called pure integrals or integrals with unit leading singularities.
\begin{figure}[t]
	\begin{center}
		\epsfxsize=15cm
		\epsffile{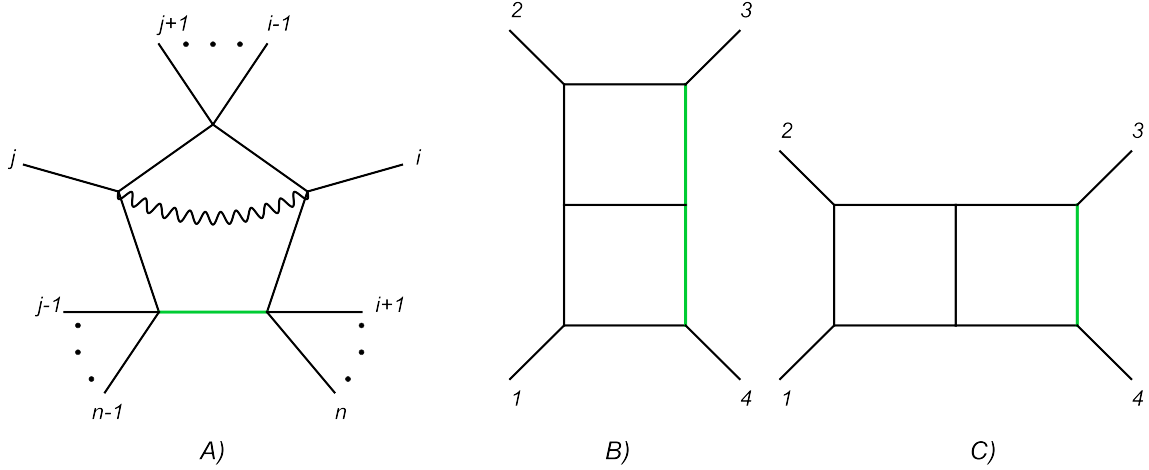}
	\end{center}\vspace{-0.2cm}
	\caption{Integrals for $A_{2,(n-2)+1}^{*(1)}(\Omega_1,\ldots,\Omega_{n-2},g^*_{n-1})$ and $A^{*(2)}_{2,2+1}(\Omega_1,\Omega_2,g_3^*)$. Green lines corresponds to eikonal propagators with shifted twistor. Wavy line corresponds to numerator of the form $\langle AB (ij)_W\rangle$, where $(ij)_W=(i-1\, i\, i+1)\cap (j-1\, j\, j+1)$.}\label{ChirlaPentagonDoubleBox}
\end{figure}

The application of the gluing operation to the on-shell integrands written in the local form follows the general rule (\ref{integrand-gluing}). Let's see some particular examples. At one-loop for MHV $n$-point integrand we have\footnote{$(i-1\, i\, i+1)\cap (j-1\, j\, j+1)\equiv Z_{i-1}Z_{i}\la i+1 j-1 j j+1 \ra+ Z_{i}Z_{i+1} \la i-1 j-1 j j+1 \ra
+Z_{i-1}Z_{i+1} \la i j-1 j j+1 \ra$} \cite{AllLoopIntegrandN4SYM,LocalIntegrands}:
\begin{equation}\label{MHVOneLoopInt}
I^{(1)}_{2,n} = \sum_{i<j} \frac{\la A B (i-1\, i\, i+1)\cap (j-1\, j\, j+1)\ra \la X i j\ra}{\la A B\, X\ra \la A B\, i-1\, i\ra \la A B\, i\, i+1\ra \la A B\, j-1\, j\ra \la A B\, j\, j+1\ra}.
\end{equation}
This expressions is cyclic invariant and sum in the above expression is independent from $X$, but contains spurious poles $\la A B\, X\ra$ term by term. If we choose $X = (k\, k+1)$ then all poles are manifestly physical but cyclic invariance will be lost. To obtain corresponding expression $I^{*(1)}_{2,(n-2) + 1}$ for the amplitude with one off-shell leg in place of two last on-shell legs $A_{2,(n-2)+1}^{*(1)}(\Omega_1,\ldots,\Omega_{n-2},g^*_{n-1})$ we just shift momentum super twistor $\ZZ_n$. Also it is convenient to choose $X = (n-1 n)$:
\begin{equation}
I^{*(1)}_{2,(n-2) + 1} = \sum_{i<j} \frac{\la A B (i-1\, i\, i+1)\cap (j-1\, j\, j+1)\ra \la n-1 n^* i j\ra}{\la A B\, n-1\, n^*\ra \la A B\, i-1\, i\ra \la A B\, i\, i+1\ra \la A B\, j-1\, j\ra \la A B\, j\, j+1\ra},
\end{equation}
where $Z^*_n$ is given by:
\begin{eqnarray}
Z^*_n = Z_n - \frac{\la p \xi\ra}{\la p 1\ra}Z_1.
\end{eqnarray}
See Fig.\ref{ChirlaPentagonDoubleBox} A. Legs $n-1$ and $n$ describe off-shell
momentum, so that $p$ and $\xi$ are light-cone vectors entering $k_T$ - decomposition of this momentum $k$.

Next, taking the expression for the integrand of 2-loop 4-point MHV on-shell amplitude \cite{AllLoopIntegrandN4SYM,LocalIntegrands}:
\begin{equation}
I_{2,4}^{(2)} = \frac{\la 2 3 4 1\ra\la 3 4 1 2\ra\la 4 1 2 3\ra}{\la A B 4 1\ra\la A B 1 2\ra\la A B 2 3\ra\la C D 2 3\ra\la C D 3 4\ra\la C D 4 1\ra\la A B C D\ra}+\mbox{cyclic, no repeat}
\end{equation}
and applying $\hat{A}_{3, 4}$ gluing operation we get for the integrand of $A^{*(2)}_{2,2+1}(\Omega_1,\Omega_2,g_3^*)$ (See Fig.\ref{ChirlaPentagonDoubleBox} B and C)
\begin{eqnarray}
I_{2,2+1}^{* (2)} &=& \frac{\la 2 3 4 1\ra\la 3 4 1 2\ra\la 4 1 2 3\ra}{\la A B 4 1\ra\la A B 1 2\ra\la A B 2 3\ra\la C D 2 3\ra\la C D 3 4^*\ra\la C D 4 1\ra\la A B C D\ra}\nonumber\\
&+&\frac{\la 3 4 1 2\ra\la 4 1 2 3\ra\la 1 2 3 4\ra}{\la A B 1 2\ra\la A B 2 3\ra\la A B 3 4^*\ra\la C D 3 4^*\ra\la C D 4 1\ra\la C D 1 2\ra\la A B C D\ra},
\end{eqnarray}
where $Z^*_4$ is given by:
\begin{eqnarray}
Z^*_4 = Z_4 - \frac{\la p \xi\ra}{\la p 1\ra}Z_1.
\end{eqnarray}

\begin{figure}[t]
	\begin{center}
		\epsfxsize=11cm
		\epsffile{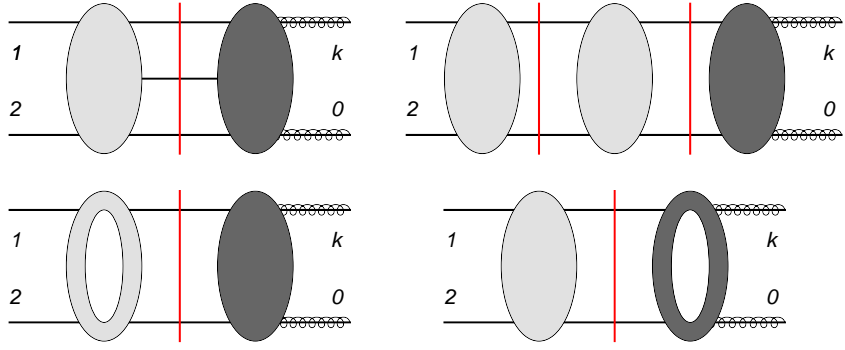}
	\end{center}\vspace{-0.2cm}
	\caption{Unitarity cuts for $A^{*(2)}_{2,2+1}(\Omega_1,\Omega_2,g_3^*)$ where only $1/l^2$ propagators have been cut. Vertical red line represents cuts of corresponding propagators. Grey blobs are on-shell amplitudes with $k=2,3$. Dark grey blobs are Willson line form factors with $k=2,3$. }\label{UnitarityCuts}
\end{figure}

As always we assume off-shell kinematics for legs 3 and 4, so that $p$ and $\xi$ are light-cone vectors entering $k_T$ - decomposition of the off-shell gluon momentum $k$.
Note also that this result is consistent with two and three particle unitarity cuts. See Fig.\ref{UnitarityCuts}.
\begin{figure}[t]
	\begin{center}
		\epsfxsize=5cm
		\epsffile{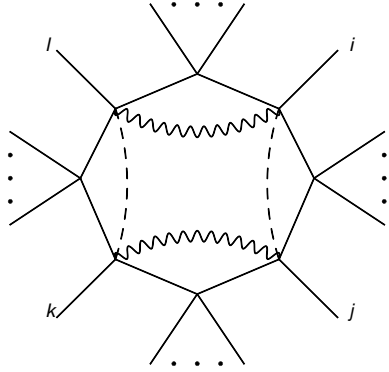}
	\end{center}\vspace{-0.2cm}
	\caption{Chiral octagons integral. Dashed line connecting $i$ and $j$ external
 legs represents numerator of the form $\langle AB ij\rangle$.}\label{ChiralOctagon}
\end{figure}

The introduced gluing operation also allows us easily obtained expressions for integrands of off-shell remainder functions starting from their on-shell counterparts. Indeed, starting from integrand for 1-loop on-shell remainder function
\begin{equation}
\mathcal{R}_{k,n}^{(1)} = I_{k,n}^{(1)} -\mathcal{P}_{n}^{4 (k-2)} I_{2,n}^{(1)}
\end{equation}
and applying gluing operation $\hat{A}_{n-1, n}$ we may obtain the expression for off-shell remainder function with one off-shell leg in place of two last on-shell legs
\begin{equation}
\mathcal{R}_{k,(n-2)+1}^{* (1)} = I_{k,(n-2)+1}^{*(1)} -\hat{A}_{n-1, n}\Big[\mathcal{P}_{n}^{4 (k-2)}\Big] I_{2,(n-2)+1}^{* (1)}\, .
\end{equation}
That is, for example taking integrand for $\mathcal{R}_{3,6}^{* (1)}$ on-shell remainder function written in terms of chiral octagons \cite{LocalIntegrands}:
\begin{multline}
\mathcal{R}_{3,6}^{*(1)} = \frac{1}{2} ([1,2,3,4,5]+[1,2,3,5,6^*]+[1,2,3,6^*,4]) I_8 (1,3,4,6^*) + \frac{1}{6}[1,2,3,4,6^*] I^{odd}_8 (1,3,4,6^*) \\
-\frac{1}{6}([1,3,4,5,6^*]-[1,2,3,4,5]) I^{odd}_8 (1,3,4,5)
+ \frac{1}{6}([1,2,4,5,6^*]+[1,3,4,5,6^*]) I^{odd}_8 (1,4,5,6^*)\, ,
\end{multline}
where
\begin{equation}
I^{odd}_8 (i,j,k,l) \equiv I_8 (i,j,k,l) - I_8 (j,k,l,i)
\end{equation}
and (see Fig. \ref{ChiralOctagon})
\begin{multline}
I_8 (i,j,k,l) =  \frac{\la A B i j\ra \la A B (j-1\, j\, j+1)\cap (k-1\, k\, k+1)\ra}{\la A B i-1\, i\ra\la A B i \, i+1\ra \la A B j-1\, j\ra \la A B j\, j+1\ra} \\
\times \frac{\la A B k l\ra\la A B (l-1\, l\, l+1)\cap (i-1\, i\, i+1)\ra}{\la A B k-1\, k\ra\la A B k\, k+1\ra\la A B l-1\, l\ra\la A B l l+1\ra}.
\end{multline}
As before $\ZZ_6^{*}$ is defined as
\begin{equation}
\ZZ_6^{*} = \ZZ_6 - \frac{\la p \xi\ra}{\la p 1\ra}\ZZ_1\, ,
\end{equation}
and we again assume off-shell kinematics for legs 5 and 6 with $p$ and $\xi$ denoting light-cone vectors entering $k_T$ - decomposition of reggeized gluon momentum.

Now we would like to show one simple but interesting test both for our tree and loop level constructions (\ref{ResultOfGluingMTwist}), (\ref{integrand-gluing}) and obtain the expression for LO BFKL kernel with gluing operation.

\subsection{LO BFKL and gluing operation}

Within BFKL approach \cite{BFKL1,BFKL2,BFKL3,BFKL4,BFKL5} amplitudes of scattering of some quantum states $A + B \to A' + B'$, which can be partons in hadron, hadrons themselves, high energy electrons etc., at large center of mass energy $\sqrt{s}$ and fixed momentum transfer $\sqrt{-t}$, $s\gg |t|$ can be represented as
\begin{equation}\label{BFKL_Ampl}
\mathcal{A}_{AB}^{A'B'} = \la \Phi_{A'A} | \e^{\alpha_s N\ln (s/s_0)~K_{BFKL} }|\Phi_{B' B}\ra\, ,
\end{equation}
where the so called impact factors  $\langle\Phi_{A'A}|$ and $|\Phi_{B'B}\rangle$ are process dependent functions and describe the transitions $A\to A'$ and $B\to B'$.
This scattering, in the mentioned above regime, can be described via interaction with special quasiparticles - so called reggeized gluons. BFKL kernel $K_{BFKL}$ describes the self interaction of these reggeized gluons. $s_0$ is some process related energy scale. See for example \cite{BalitskyLectures} for detailed discussion.

Let us now calculate the LO kernel of BFKL equation in $\mathcal{N}=4$ SYM with the use of our gluing operation.  At LO order it is
given by two contribution so called real and virtual one. Consider virtual contribution first (also see fig. \ref{BFKL_F_Diagrams_LO} A).

\subsubsection{Virtual part of LO BFKL}

To compute virtual contribution to the LO BFKL we need the Regge trajectory. The latter could be conveniently extracted from the one-loop correlation function of two Wilson lines playing the role of sources for reggeized gluons \cite{BalitskyLectures}. Namely, we have to compute the following off-shell amplitude:
\begin{equation}
\langle0|\mathcal{W}_{p_1}(k)\mathcal{W}_{p_2}(-k)|0\rangle=A_{2, 0+2}^{*}(g_1^*,g_2^*) = \hat{A}_{12}\circ \hat{A}_{34} \left[
A_{2,4} (1^{-}, 2^{+}, 3^{-}, 4^{+})
\right]\, .
\end{equation}
At tree level we have\footnote{The gluing details are similar to those considered in sections \ref{ReggeonsWilsonLines} and \ref{offshellBCFW}. }
\begin{equation}
A_{2, 0+2}^{* (0)} = \frac{\abr{p_1\, \xi_1}}{\kappa_1^*} \frac{\abr{p_2\, \xi_2}}{\kappa_2^*}\left(
\frac{\abr{1\, 3}^4}{\abr{1\, 2}\abr{2\, 3}\abr{3\, 4}\abr{4\, 1}}
\right)\Bigg|_{*} \prod_{i=1}^2 \frac{1}{\beta_{1,(i)}^2\beta_{2,(i)}} \frac{d\beta_{1, (i)}\wedge d\beta_{2, (i)}}{\beta_{1, (i)}\, \beta_{2, (i)}}\, ,
\end{equation}
where $\beta_{1,(1)}, \beta_{2,(1)}$ parameters correspond to $\hat{A}_{12}$ gluing operation and those with $(2)$ subscripts to $\hat{A}_{34}$.  Evaluating $\Big|_{*}$ substitutions and taking composite residues at $\beta_{1, (i)} = -1$, $\beta_{2, (i)} = 0$ we get
\begin{equation}
A_{2, 0+2}^{* (0)} = -\frac{\abr{p_1\, p_2}^2}{\kappa_1^*\kappa_2^*}\, .
\end{equation}
Now we should recall that the Wilson lines were used here to describe  scattering of two fast moving particles at high energy\footnote{For the introduction to corresponding description see \cite{BalitskyLectures}.}. This restricts further our kinematics, so that $p_1\cdot p_2 = s/2$ ($s$ is the usual Mandelstam variable) and momentum transfer between two particles is restricted by two orthogonality conditions $k\cdot p_1 = k\cdot p_2 = 0$. The latter two conditions allow us to write down transverse momentum transfer as
\begin{equation}
k = c_1 \vll_{p_1} \vlt_{p_2} + c_2 \vll_{p_2} \vlt_{p_1}\, ,
\end{equation}
so that $t \equiv k^2 = c_1 c_2 s$ and\footnote{It is convenient here to chose $\xi_1 = p_2$ and $\xi_2 = p_1$.}
\begin{equation}
\kappa_1^* \kappa_2^* = \frac{c_2 \abr{p_1\, p_2}\sbr{p_1\, p_2}}{\sbr{p_1\, p_2}} \frac{c_1 \abr{p_2\, p_1}\sbr{p_2\, p_1}}{\sbr{p_2\, p_1}} = - c_1 c_2 \abr{p_1\, p_2}^2 = -\frac{t}{s}  \abr{p_1\, p_2}^2.
\end{equation}
Then for $A_{2, 0+2}^{* (0)}$ amplitude we have
\begin{equation}
A_{2, 0+2}^{* (0)} = \frac{s}{t}\, .
\end{equation}
Now let us turn to the integrand of the corresponding one-loop amplitude. The latter is given for $n=4$ by (\ref{MHVOneLoopInt}):
\begin{eqnarray}
	I_{2,0+2}^{* (1)} =
	\frac{\la 1 2 3 4\ra^2}{\la 1 2^* AB\ra
	\la 23 AB\ra \la 3 4^* AB\ra \la 41 AB\ra },
\end{eqnarray}
with
\begin{eqnarray}
	Z^*_2=Z_2-\frac{\la p_1\xi_1 \ra}{\la p_1p_2\ra}Z_3, ~
	Z^*_4=Z_4-\frac{\la p_2\xi_2 \ra}{\la p_2p_1\ra}Z_1.
\end{eqnarray}
This expression can be rewritten in spinor helicity variables as:
\begin{equation}
I_{2,0+2}^{* (1)} = \frac{t^2 \abr{p_1\, p_2}^2}{4\kappa_1^* \kappa_2^*} \frac{1}{l^2 (l+k)^2 ~l\cdot p_1~ l\cdot p_2} =   \frac{s t}{4} \frac{1}{l^2 (l+k)^2~ l\cdot p_1~ l\cdot p_2}\, .
\end{equation}
The same result can also be obtained within helicity spinor picture
\begin{equation}
I_{2,0+2}^{* (1)} =  \hat{A}_{12}\circ \hat{A}_{34} \left[
I_{2,4}^{(1)} (1^{-}, 2^{+}, 3^{-}, 4^{+})
\right]\, ,
\end{equation}
where we arrange loop momenta as:
\begin{equation}
I_{2,4}^{(1)} (1^{-}, 2^{+}, 3^{-}, 4^{+}) =  \frac{(q_1+q_2)^2 (q_2+q_3)^2}{l^2 (l+q_2)^2 (l+q_1+q_2)^2 (l-q_3)^2}\, .
\end{equation}
In the expression above $q_i, \, (i=1,\ldots , 4)$ are momenta of external gluons and $l$ is loop momentum.

In LO BFKL regime we are interested in leading logarithmic approximation (LLA) to high-energy scattering amplitude. The latter could be obtained using Sudakov decomposition of loop integration momentum and retaining only logarithmic in Mandelstam invariant $s$ contribution. That is
\begin{eqnarray}
l=\alpha p_1+\beta p_2 +l_{\perp},\quad p_i\cdot l_{\perp} = 0,\quad k\cdot p_i = 0,
\end{eqnarray}
\begin{eqnarray}
d^Dl=\frac{s}{2}~d\alpha ~d\beta ~d^{D-2}l_{\perp}.
\end{eqnarray}
and we are interested in the following regime (here $m$ is some problem related mass scale):
\begin{eqnarray}
1 \gg \alpha \gg \beta\sim\frac{m^2}{s} \ll 1.
\end{eqnarray}
Then
\begin{eqnarray}
p_2\cdot l &=& \alpha s/2 + p_1\cdot l_{\perp} = \alpha s/2 \\
p_1\cdot l &=& \beta s/2 + p_2\cdot l_{\perp} = \beta s/2 \\
l^2 &=& \alpha\beta s/2 - l_{\perp}^2 \\
(l+k)^2 &=& \alpha\beta s/2 - (l_{\perp} + k_{\perp})^2.
\end{eqnarray}
Now taking residue in $\beta$ at $0$ and integrating over $\alpha$ from $m^2/s$ to $1$ we get
\begin{multline}
\int \frac{d^4l}{(2\pi)^4}\frac{1}{l^2(l+k)^2(p_1l)(p_2l)}
= \frac{s}{2 (2\pi )^4}\int \frac{d\alpha}{\alpha s/2}
\frac{d\beta}{\beta s/2}\frac{d^{D-2}l_{\perp}}{[\alpha\beta s/2 - l_{\perp}^2][\alpha\beta s/2 - (l_{\perp} + k_{\perp})^2]} \\
= \frac{1}{4 \pi^3 s}\log \left(\frac{s}{m^2}\right)\int \frac{d^2 l_{\perp}}{l_{\perp}^2 (l_{\perp} + k_{\perp})^2}.
\end{multline}
So finally we get
\begin{equation}\label{BFKL_Virt}
A_{2, 0+2}^{* (0+1)} = A_{2, 0+2}^{* (0)} \left\{
1 - \frac{g^2}{16\pi^3} \log \left(\frac{s}{m^2}\right)
 \int \frac{k_{\perp}^2 d^2 l_{\perp}}{l_{\perp}^2 (l_{\perp} + k_{\perp})^2} \right\}\, .
\end{equation}
This expression tells us that in LLA approximation\footnote{We should resum logarithmic terms, which exponentiate} with account for color factor ($C_A = N$ for $SU(N)$ gauge group) for reggeized gluon propagator we get
\begin{equation}
\langle0|\mathcal{W}_{p_1}(k)\mathcal{W}_{p_2}(-k)|0\rangle\Big{|}_{LLA}\sim\frac{1}{k_{\perp}^2} \left(
\frac{s}{m^2}
\right)^{\omega (k_{\perp}^2)}\, ,
\end{equation}
where\footnote{Here we introduced dimensional regularization of otherwise divergent integral.} ($\alpha_s = g^2/(4\pi)\; ,\, t = -k_{\perp}^2$):
\begin{equation}
\omega (t) = - \alpha_s N \int \frac{d^{2+\ep} l_{\perp}}{(2\pi)^{2+\ep}} \frac{k_{\perp}^2 }{l_{\perp}^2 (l_{\perp} + k_{\perp})^2} =  - \frac{N\alpha_s }{4\pi} \frac{2 (k_{\perp}^2)^{\ep/2}}{\ep}
\end{equation}
is the famous LO BFKL Regge trajectory, which at LO is the same in QCD and $\mathcal{N}=4$ SYM.
See for example \cite{NLOBFKL-QCD-SUSY,DGLAP-BFKL-N4SYM1,DGLAP-BFKL-N4SYM2}. Using this result for
virtual part of BFKL kernel we can write \cite{BalitskyLectures}:
\begin{equation}
- \alpha_s N~K_{BFKL}^V=-\frac{1}{2}\delta^{(2)}(k-k')\left(\omega (k_{\perp})+\omega (k_{\perp}-r_{\perp})\right).
\end{equation}
Here $r$ is the momentum transfer for $A\to A'$ scattering $r=p_{A'}-p_A$.

In conclusion we would like also to note the following interesting fact. In $\mathcal{N}=4$ SYM the four point on-shell amplitude $A_{2,4}$ has a remarkable property\footnote{See the discussion in \cite{ConformalProperties4point}.}
of being Regge exact, i.e. the contribution of the gluon Regge trajectory to the amplitude ($c(t)$ is the gluon impact factor)
\begin{equation}
A_{2,4}(s,t) = c(t)^2 \left(\frac{s}{-t}\right)^{\omega (t)} +\; \text{subleading terms in}\; \frac{|t|}{s}\, .
\end{equation}
coincides with the exact expression for $A_{2,4}(s,t)$ as a function of arbitrary $s$ and $t$.

\subsubsection{Real part of LO BFKL}
\begin{figure}[t]
	\begin{center}
		\epsfxsize=9cm
		\epsffile{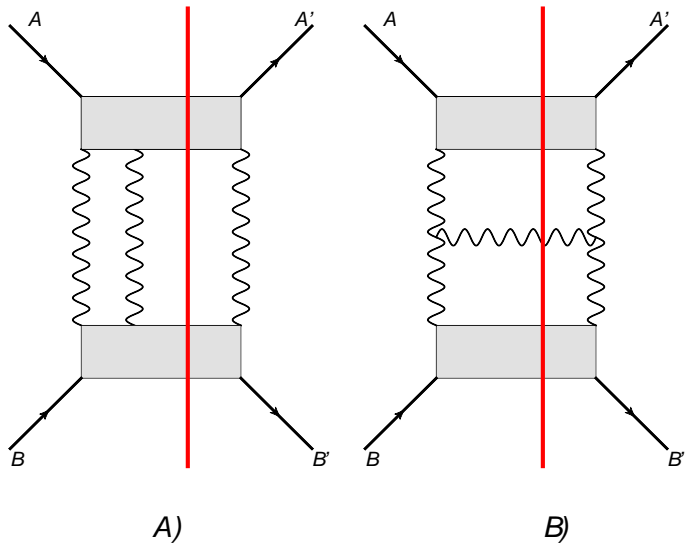}
	\end{center}\vspace{-0.2cm}
	\caption{Typical Feynman diagrams contributing to (\ref{BFKL_Ampl}) i.e. to the BFKL kernel at LO in $\mathcal{N}=4$ SYM. At large center of mass energy $\sqrt{s}$ and fixed momentum transfer $\sqrt{-t}$ the asymptotical behaviour of an amplitude is given by its imaginary part \cite{BalitskyLectures,LipatovEL2}: $\mathcal{A}_{AB}^{A'B'}|_{s \gg 1 }\sim \mbox{Im}[\mathcal{A}_{AB}^{A'B'}]$. Grey squares represents impact factors, wavy lines
represents gluon propagators, vertical red line represents cuts of corresponding propagators and impact factors. Diagrams of type A) gives contribution to $K^V_{BFKL}$ and their total sum is equivalent to evaluation of (\ref{BFKL_Virt}). Diagrams of type B) give contribution to $K_{BFKL}^R$ and their total sum is equivalent to evaluation of (\ref{BFKL_Real_Rad}).}\label{BFKL_F_Diagrams_LO}
\end{figure}

Now let's consider real contribution. This contribution is given by the integrated product of two, so called, Lipatov's $L_{\mu}$ RRP vertexes \cite{BalitskyLectures}, see fig. \ref{BFKL_F_Diagrams_LO} B. To compute this contribution we may note, that Lipatov's RRP $L_{\mu}$ vertex (tree level reggeon-reggeon-particle amplitude) is related to reggeon amplitudes $A^*_{2,2+1}(g_1^*,g_2^*,3^+)$ and $A^*_{3,2+1}(g_1^*,g_2^*,3^-)$ as
\begin{equation}
L_{\mu} (k,k') =  (k' + k)_{\mu} + n_{\mu}^{-} \left(
\frac{k^2}{k'^{-}} - k^{+}
\right) + n_{\mu}^{+} \left(
\frac{k'^{2}}{k^{+}} - k'^{-}
\right),
\end{equation}
\begin{eqnarray}
	A^*_{2,2+1}(g_1^*,g_2^*,3^+)&=&\delta^4(k-k'-q_3)\bold{A}^*_{2,2+1}(k,-k',-q_3)=
	\delta^4(k-k'-q_3)\frac{\epsilon_3^{\mu,+}L_{\mu} (k,k')}{k^2k^{'2}},\nonumber\\
  A^*_{3,2+1}(g_1^*,g_2^*,3^-)&=&\delta^4(k-k'-q_3)\bold{A}^*_{3,2+1}(k,-k',-q_3)=
	\delta^4(k-k'-q_3)\frac{\epsilon_3^{\mu,-}L_{\mu} (k,k')}{k^2k^{'2}},\nonumber\\
\end{eqnarray}
which in their turns could be obtained with two our gluing operations applied to 5-point on-shell amplitude $A_{2,5}$. Here $k$, $k'$ are reggeized gluons $g_1^*$ and $g_1^*$ momenta with $k-k'-q_3=0$ and
$n^{\pm}$ are normalized light like directions for reggeized gluons
\begin{equation}
n^{-} = \frac{2 p_1}{\sqrt{s}}~, \quad n^{+} = \frac{2 p_2}{\sqrt{s}},~ (n^{-}n^{+})=2,~(kn^{\pm})\equiv
k^{\pm},
\end{equation}
and $\epsilon^{\pm}_3$ are polarization vectors of on-shell gluon with momentum $-q_3$. It is assumed that in the definitions of $A^*_{2,2+1}(g_1^*,g_2^*,3^+)$ and $A^*_{3,2+1}(g_1^*,g_2^*,3^-)$ amplitudes one has to
take in $k_T$ decomposition of $k$ and $k'$ momenta direction vectors as $p_1=n^{-}$ and $p_2=n^{+}$. We have also defined functions $\bold{A}^*_{2,2+1}$ and $\bold{A}^*_{3,2+1}$
which are given by corresponding Wilson line form factors stripped from
momentum conservation delta functions:
\begin{eqnarray}
	\bold{A}^*_{2,2+1}(k_1,k_2,q_3)&=&\frac{1}{\kappa_1^*\kappa_2^*}
	\frac{\langle n^{-}n^{+} \rangle^3}{\langle 3n^{+} \rangle\langle n^{-}3 \rangle},
	\nonumber\\
  \bold{A}^*_{3,2+1}(k_1,k_2,q_3)&=&\frac{1}{\kappa_1\kappa_2}\frac{[n^{-}n^{+}]^3}{[3n^{+}][n^{-}3]}.
\end{eqnarray}

Performing Sudakov decomposition\footnote{See for example \cite{BalitskyLectures} for details.} of reggeized gluon momentum the contribution of real radiation to BFKL kernel takes the form \cite{BalitskyLectures}:
\begin{equation}\label{BFKL_Real_Rad}
K^R_{BFKL} (k_{\perp}, k'_{\perp}, r) \ln\frac{s}{m^2} =  \frac{s}{2}\int d\alpha_{k} d\beta_{k'} L^{\mu} (k, k') L_{\mu} (r-k, r-k')  \frac{\delta (\alpha_{k}\beta_{k'} s + (k_{\perp} - k'_{\perp})^2)}{k_{\perp}^{'2} (r_{\perp} - k'_{\perp})^2} \, ,
\end{equation}
Note that factor $L^{\mu} (k, k') L_{\mu} (r-k, r-k')=g_{\mu\nu}L^{\mu} (k, k') L^{\nu} (r-k, r-k')$
can be rewritten purely in terms of Wilson line form factors.
Namely using gauge invariance of $A^*_{2,2+1}(g_1^*,g_2^*,3^+)$ and $A^*_{3,2+1}(g_1^*,g_2^*,3^-)$ we can replace
\begin{eqnarray}
	g_{\mu\nu}\mapsto \sum_{i=\pm}\epsilon_{\mu}^{(i)}\epsilon_{\nu}^{(i)},
\end{eqnarray}
so that ($r-k\equiv m$ and $r-k'\equiv m'$)
\begin{eqnarray}
	\frac{L^{\mu} (k, k') L_{\mu} (m, m')}{k^2k^{'2}m^2m^{'2}}&=&
	\bold{A}^*_{2,2+1}(k,-k',-q_3)\bold{A}^*_{2,2+1}(m,-m',-q_3)\nonumber\\
	&+&
	\bold{A}^*_{3,2+1}(k,-k',-q_3)\bold{A}^*_{3,2+1}(m,-m',-q_3).
\end{eqnarray}
Also note that in this case no other particles besides gluons from $\mathcal{N}=4$ SYM supermultiplet give contribution to real radiation. This happens due to the R-charge conservation.

The integral over $\beta_{k'}$ is taken with the help of $\delta$ - function, while the integration over $\alpha_k$ is performed over the interval $[\frac{m^2}{s}, 1]$. This way we get
\begin{equation}
K^R_{BFKL} (k_{\perp}, k'_{\perp}, r) = -\frac{r_{\perp}^2}{k_{\perp}^{' 2} (r_{\perp} - k'_{\perp})^2}  + \frac{k_{\perp}^2}{k_{\perp}^{' 2} (k_{\perp} - k'_{\perp})^2} + \frac{(r_{\perp} - k_{\perp})^2}{(r_{\perp} - k'_{\perp})^2 (k_{\perp} - k'_{\perp})^2}.
\end{equation}
Altogether with account for the Regge trajectories contributions we recover LO expression for BFKL kernel $K_{BFKL}=K_{BFKL}^R+K_{BFKL}^V$ \cite{BalitskyLectures}:
\begin{multline}
K_{BFKL} (k_{\perp}, k'_{\perp}, r)  = -\frac{r_{\perp}^2}{k_{\perp}^{' 2} (r_{\perp} - k'_{\perp})^2}  + \frac{k_{\perp}^2}{k_{\perp}^{' 2} (k_{\perp} - k'_{\perp})^2} + \frac{(r_{\perp} - k_{\perp})^2}{(r_{\perp} - k'_{\perp})^2 (k_{\perp} - k'_{\perp})^2} \\
- \frac{1}{2}\delta^{(2)} (k-k') \int\frac{d^2 l_{\perp}}{4\pi^2}\left\{
\frac{k_{\perp}^2}{l_{\perp}^2 (k_{\perp} - l_{\perp})^2} + \frac{(k_{\perp} - r_{\perp})^2}{(l_{\perp} - r_{\perp})^2 (k_{\perp} - l_{\perp})^2}
\right\}\, .
\end{multline}

\section{Conclusion}

In this paper we considered the derivation of the BCFW recurrence relation for the Wilson line form factors and correlation functions (off-shell reggeon amplitudes) both at tree and at integrand level. We have shown that starting from the BCFW recursion for on-shell amplitudes and using so
called ``gluing operator'' one can obtain recursion relations for the Wilson line form factors. The latter is true both at tree and integrand level in helicity spinor and momentum twistor representations. The gluing operation also allows one easily convert known local integrands of the on-shell amplitudes into integrands of the Wilson line form factors. These results are condensed in formulas (\ref{ResultOfGluingMTwist}), (\ref{ResultOfGluingMTwistAmpl}) and (\ref{BCFWInLoopsWLFF}), (\ref{integrand-gluing}) for tree and loop level correspondingly. We have verified our considerations by reproducing LO BFKL kernel. We also made some predictions regarding the structure of the integrands of Wilson line form factors at higher loops/large number of external states.

As far as we can understand our construction is not limited to the Wilson line operators only. Indeed, using \cite{ambitwistorFormfactors} similar gluing operator the form factors of
stress tensor operator supermultiplet could be constructed. Also, presumably, analogs of gluing operator for all other type of local operators in $\mathcal{N}=4$ SYM theory should exist. The only real obstacle in this direction is that at the level of integrands for local single trace operators we should account for nonplanar contributions. So the notion of ``integrand'' in this case is somewhat obscure at first sight. Nevertheless we think that one can still introduce integrands in setup similar to considerations in \cite{Du:2014jwa}, where nonplanar contributions to the on-shell amplitudes were considered in momentum twistor variables.

We hope that the presented results will be interesting and useful for people both from $\mathcal{N}=4$ SYM ``amplitudology'' and BFKL/reggeon physics communities.

\section*{Acknowledgements}

This work was supported by RFBR grants \# 17-02-00872, \# 16-02-00943, contract \# 02.A03.21.0003 from 27.08.2013 with Russian Ministry of Science and Education and Heisenberg-Landau program. The work of L.V. Bork was supported by the grant  \# 17-13-325-1  of the “Basis” foundation for theoretical physics.

\appendix

\section{Gluing operation and Grassmannians }\label{appA}
Let's see how the use of the gluing operation in momentum twistors could easily reproduce known Grassmannian integral representations for the tree-level off-shell amplitudes \cite{offshell-1leg,offshell-multiple}. We start with the Grassmannian integral representation for the on-shell amplitudes in momentum twistors:
\begin{multline}
\mathcal{L}_{k,n+2} = \frac{A_{k,n+2}}{A_{2,n+2}}
 = \int
 \frac{d^{(k-2)\times (n+2)}D}{\text{Vol}[GL(k-2)]}
 \frac{\delta^{ 4 (k-2) | 4 (k-2)} (D\cdot \mathcal{Z})}{(1 \; \ldots \; k-2) \;\ldots \; (n+2 \; \ldots \; k-3)}\, ,
\end{multline}
here (also similar notations are used in (\ref{GluingGrassmannianIntegrals}))
\begin{eqnarray}
	\delta^{ 4 (k-2) | 4 (k-2)} (D\cdot \mathcal{Z})=\prod_{a=1}^{k-2}\delta^{4|4} \left(\sum_{i=1}^{{n+2}}D_{ai}\ZZ_i\right),
\end{eqnarray}
and $(i_1,\ldots,i_{k-2})$ is minor constructed from columns of $D$ matrix with numbers $i_1,\ldots,i_{k-2}$.
Applying to this expression the gluing operation $\hat{A}_{n+1,n+2}$ amounts to the following shifts of momentum super twistors:
\begin{align}
&\ZZ_1 \to  (1+\alpha_1\alpha_2)\ZZ_1 + \alpha_1 \ZZ_{n+2} \equiv \ZZ_1'\, , \\
&\ZZ_{n+2} \to \ZZ_{n+2} + \alpha_2 \ZZ_1 \equiv \ZZ_{n+2}'\, ,
\end{align}
so that
\begin{equation}
D\cdot \ZZ \to D'\cdot \ZZ \, ,
\end{equation}
where
\begin{align}
&D'_1 = (1+\alpha_1 \alpha_2) D_1 + \alpha_2 D_{n+2}\, , \\
&D'_{n+2} = \alpha_1 D_1 + D_{n+2}
\end{align}
All other momentum super twistors are unshifted and we have $D'_i = D_i$.
The inverse transformation from $D$'s to $D'$'s is then given by
\begin{align}
&D_1 = D'_1 - \alpha_2 D'_{n+2}\, , \\
&D_{n+2} = -\alpha_1 D'_1 + (1+\alpha_1\alpha_2) D'_{n+2}\, .
\end{align}
With these transformations it is easy to write down transformation rules for minors. For example, we have
\begin{align}
&(1\ldots k-2) \to (1\ldots k-2)' - \alpha_2 (n+2\, 2\ldots k-2)' \\
&(n-k+5\ldots n+2)\to -\alpha_1 (n-k+5\ldots n+1\, 1)' + (1+\alpha_1\alpha_2) (n-k+5\ldots n+2)'
\end{align}
Finally performing transition in the Grassmannian integral from $D$'s to $D'$'s and taking residues at  $\alpha_1 = 0$ and $\alpha_2 = - \frac{\la n+1\, n+2\ra}{\la n+1\, 1\ra} = -\frac{\la p_{n+1}\, \xi_{n+1}\ra}{\la p_{n+1}\, 1\ra}$ we  get
\begin{eqnarray}
\frac{A^{*}_{k,n+1}}{A^{*}_{2,n+1}} = \int_{\Gamma}
\frac{d^{(k-2)\times (n+2)}D'}{\text{Vol}[GL(k-2)]}\frac{1}{1+\frac{\la p_{n+1}\, \xi_{n+1}\ra}{\la p_{n+1}\, 1\ra}\frac{(n+2 \; 2 \; \ldots \; k-2)'}{(1 \; \ldots \; k-2)'}}
\frac{\delta^{ 4 (k-2) | 4 (k-2)} (D'\cdot \mathcal{Z})}{(1 \; \ldots \; k-2)' \;\ldots \; (n+2 \; \ldots \; k-3)'}\, , \nonumber\\
\end{eqnarray}
Similarly applying several gluing operations we recover formula \cite{offshell-multiple}.

\bibliographystyle{ieeetr}
\bibliography{refs}

\begin{thebibliography}{10}

\bibitem{Henrietta_Amplitudes}
H.~Elvang and Y.-t. Huang, ``{Scattering Amplitudes},'' 2013.

\bibitem{Talesof1001Gluons}
S.~Weinzierl, ``{Tales of 1001 Gluons},'' 2016.

\bibitem{FormFactorsPeriodicWilsonLoops}
A.~Brandhuber, B.~Spence, G.~Travaglini, and G.~Yang, ``{Form Factors in N=4
  Super Yang-Mills and Periodic Wilson Loops},'' {\em JHEP}, vol.~01, p.~134,
  2011.

\bibitem{HarmonyofFF_Brandhuber}
A.~Brandhuber, O.~Gurdogan, R.~Mooney, G.~Travaglini, and G.~Yang, ``{Harmony
  of Super Form Factors},'' {\em JHEP}, vol.~10, p.~046, 2011.

\bibitem{Engelund:2012re}
O.~T. Engelund and R.~Roiban, ``{Correlation functions of local composite
  operators from generalized unitarity},'' {\em JHEP}, vol.~03, p.~172, 2013.

\bibitem{PolytopesFormFactors}
L.~V. Bork, ``{On form factors in $ \mathcal{N}=4 $ SYM theory and
  polytopes},'' {\em JHEP}, vol.~12, p.~111, 2014.

\bibitem{SuperFormFactorsHalfBPS}
B.~Penante, B.~Spence, G.~Travaglini, and C.~Wen, ``{On super form factors of
  half-BPS operators in N=4 super Yang-Mills},'' {\em JHEP}, vol.~04, p.~083,
  2014.

\bibitem{FormFactorsGrassmanians}
R.~Frassek, D.~Meidinger, D.~Nandan, and M.~Wilhelm, ``{On-shell diagrams,
  Graßmannians and integrability for form factors},'' {\em JHEP}, vol.~01,
  p.~182, 2016.

\bibitem{SoftTheoremsFormFactors}
L.~V. Bork and A.~I. Onishchenko, ``{On soft theorems and form factors in $
  \mathcal{N}=4 $ SYM theory},'' {\em JHEP}, vol.~12, p.~030, 2015.

\bibitem{TwistorFormFactors1}
L.~Koster, V.~Mitev, M.~Staudacher, and M.~Wilhelm, ``{Composite Operators in
  the Twistor Formulation of N=4 Supersymmetric Yang-Mills Theory},'' {\em
  Phys. Rev. Lett.}, vol.~117, no.~1, p.~011601, 2016.

\bibitem{TwistorFormFactors2}
L.~Koster, V.~Mitev, M.~Staudacher, and M.~Wilhelm, ``{All tree-level MHV form
  factors in $ \mathcal{N} $ = 4 SYM from twistor space},'' {\em JHEP},
  vol.~06, p.~162, 2016.

\bibitem{TwistorFormFactors3}
L.~Koster, V.~Mitev, M.~Staudacher, and M.~Wilhelm, ``{On Form Factors and
  Correlation Functions in Twistor Space},'' 2016.

\bibitem{GeneralizedUnitarityFormFactors}
L.~V. Bork, ``{On NMHV form factors in N=4 SYM theory from generalized
  unitarity},'' {\em JHEP}, vol.~01, p.~049, 2013.

\bibitem{BrandhuberConnectedPrescription}
A.~Brandhuber, E.~Hughes, R.~Panerai, B.~Spence, and G.~Travaglini, ``{The
  connected prescription for form factors in twistor space},'' {\em JHEP},
  vol.~11, p.~143, 2016.

\bibitem{HeConnectedFormulaFormFactors}
S.~He and Z.~Liu, ``{A note on connected formula for form factors},'' {\em
  JHEP}, vol.~12, p.~006, 2016.

\bibitem{DilatationOperatorFormFactors5}
A.~Brandhuber, B.~Penante, G.~Travaglini, and C.~Wen, ``{The last of the simple
  remainders},'' {\em JHEP}, vol.~08, p.~100, 2014.

\bibitem{FormFactorsYsystem1}
J.~Maldacena and A.~Zhiboedov, ``{Form factors at strong coupling via a
  Y-system},'' {\em JHEP}, vol.~11, p.~104, 2010.

\bibitem{FormFactorsYsystem2}
Z.~Gao and G.~Yang, ``{Y-system for form factors at strong coupling in $AdS_5$
  and with multi-operator insertions in $AdS_3$},'' {\em JHEP}, vol.~06,
  p.~105, 2013.

\bibitem{offshell-1leg}
L.~V. Bork and A.~I. Onishchenko, ``{Wilson lines, Grassmannians and Gauge
  Invariant Off-shell Amplitudes in N=4 SYM},'' 2016.

\bibitem{offshell-multiple}
L.~V. Bork and A.~I. Onishchenko, ``{Grassmannian Integral for General Gauge
  Invariant Off-shell Amplitudes in N=4 SYM},'' 2016.

\bibitem{BKV_SuperForm}
L.~V. Bork, D.~I. Kazakov, and G.~S. Vartanov, ``{On MHV Form Factors in
  Superspace for N=4 SYM Theory},'' {\em JHEP}, vol.~10, p.~133, 2011.

\bibitem{KotkoWilsonLines}
P.~Kotko, ``{Wilson lines and gauge invariant off-shell amplitudes},'' {\em
  JHEP}, vol.~07, p.~128, 2014.

\bibitem{q2zeroFormFactors}
L.~V. Bork and A.~I. Onishchenko, ``{Grassmannians and form factors with
  q$^{2}$ = 0 in $ \mathcal{N}=4$ SYM theory},'' {\em JHEP}, vol.~12, p.~076,
  2016.

\bibitem{ambitwistorFormfactors}
L.~V. Bork and A.~I. Onishchenko, ``{Four dimensional ambitwistor strings and
  form factors of local and Wilson line operators},'' 2017.

\bibitem{Eden:2017fow}
B.~Eden, P.~Heslop, and L.~Mason, ``{The Correlahedron},'' {\em JHEP}, vol.~09,
  p.~156, 2017.

\bibitem{Amplituhdron_1}
N.~Arkani-Hamed and J.~Trnka, ``{The Amplituhedron},'' {\em JHEP}, vol.~10,
  p.~030, 2014.

\bibitem{Amplituhdron_2}
N.~Arkani-Hamed and J.~Trnka, ``{Into the Amplituhedron},'' {\em JHEP},
  vol.~12, p.~182, 2014.

\bibitem{Amplituhdron_3}
Y.~Bai and S.~He, ``{The Amplituhedron from Momentum Twistor Diagrams},'' {\em
  JHEP}, vol.~02, p.~065, 2015.

\bibitem{Amplituhdron_7}
N.~Arkani-Hamed, Y.~Bai, and T.~Lam, ``{Positive Geometries and Canonical
  Forms},'' 2017.

\bibitem{Amplituhdron_8}
N.~Arkani-Hamed, H.~Thomas, and J.~Trnka, ``{Unwinding the Amplituhedron in
  Binary},'' {\em JHEP}, vol.~01, p.~016, 2018.

\bibitem{Bork2010FormFact}
L.~V. Bork, D.~I. Kazakov, and G.~S. Vartanov, ``{On form factors in N=4
  sym},'' {\em JHEP}, vol.~02, p.~063, 2011.

\bibitem{2loopFormFactors2012}
A.~Brandhuber, G.~Travaglini, and G.~Yang, ``{Analytic two-loop form factors in
  N=4 SYM},'' {\em JHEP}, vol.~05, p.~082, 2012.

\bibitem{DilatationOperatorFormFactors1}
M.~Wilhelm, ``{Amplitudes, Form Factors and the Dilatation Operator in
  $\mathcal{N}=4$ SYM Theory},'' {\em JHEP}, vol.~02, p.~149, 2015.

\bibitem{DilatationOperatorFormFactors2}
D.~Nandan, C.~Sieg, M.~Wilhelm, and G.~Yang, ``{Cutting through form factors
  and cross sections of non-protected operators in $ \mathcal{N}=4 $ SYM},''
  {\em JHEP}, vol.~06, p.~156, 2015.

\bibitem{DilatationOperatorFormFactors3}
F.~Loebbert, D.~Nandan, C.~Sieg, M.~Wilhelm, and G.~Yang, ``{On-Shell Methods
  for the Two-Loop Dilatation Operator and Finite Remainders},'' {\em JHEP},
  vol.~10, p.~012, 2015.

\bibitem{DilatationOperatorFormFactors4}
F.~Loebbert, C.~Sieg, M.~Wilhelm, and G.~Yang, ``{Two-Loop SL(2) Form Factors
  and Maximal Transcendentality},'' {\em JHEP}, vol.~12, p.~090, 2016.

\bibitem{FormFactorsColorKinematic}
R.~H. Boels, B.~A. Kniehl, O.~V. Tarasov, and G.~Yang, ``{Color-kinematic
  Duality for Form Factors},'' {\em JHEP}, vol.~02, p.~063, 2013.

\bibitem{FormFactorsColorKinematic5loop}
G.~Yang, ``{Color-kinematics duality and Sudakov form factor at five loops for
  N=4 supersymmetric Yang-Mills theory},'' {\em Phys. Rev. Lett.}, vol.~117,
  no.~27, p.~271602, 2016.

\bibitem{HennGehrmann_FormF3loops_2011}
T.~Gehrmann, J.~M. Henn, and T.~Huber, ``{The three-loop form factor in N=4
  super Yang-Mills},'' {\em JHEP}, vol.~03, p.~101, 2012.

\bibitem{Boels:2015yna}
R.~Boels, B.~A. Kniehl, and G.~Yang, ``{Master integrals for the four-loop
  Sudakov form factor},'' {\em Nucl. Phys.}, vol.~B902, pp.~387--414, 2016.

\bibitem{Banerjee:2016kri}
P.~Banerjee, P.~K. Dhani, M.~Mahakhud, V.~Ravindran, and S.~Seth, ``{Finite
  remainders of the Konishi at two loops in $ \mathcal{N}=4 $ SYM},'' {\em
  JHEP}, vol.~05, p.~085, 2017.

\bibitem{DilatationOperatorFormFactors6}
A.~Brandhuber, M.~Kostacinska, B.~Penante, G.~Travaglini, and D.~Young, ``{The
  SU(2|3) dynamic two-loop form factors},'' {\em JHEP}, vol.~08, p.~134, 2016.

\bibitem{Huang2016FormFactorBoundaryContribution}
R.~Huang, Q.~Jin, and B.~Feng, ``{Form Factor and Boundary Contribution of
  Amplitude},'' {\em JHEP}, vol.~06, p.~072, 2016.

\bibitem{Engelund:2015cfa}
O.~T. Engelund, ``{Lagrangian Insertion in the Light-Like Limit and the
  Super-Correlators/Super-Amplitudes Duality},'' {\em JHEP}, vol.~02, p.~030,
  2016.

\bibitem{Kotko:2017nkx}
P.~Kotko and A.~M. Stasto, ``{Wilson lines in the MHV action},'' {\em JHEP},
  vol.~09, p.~047, 2017.

\bibitem{vanHamerenBCFW1}
A.~van Hameren, ``{BCFW recursion for off-shell gluons},'' {\em JHEP}, vol.~07,
  p.~138, 2014.

\bibitem{vanHamerenBCFW2}
A.~van Hameren and M.~Serino, ``{BCFW recursion for TMD parton scattering},''
  {\em JHEP}, vol.~07, p.~010, 2015.

\bibitem{vanHamerenBCFW3}
K.~Kutak, A.~Hameren, and M.~Serino, ``{QCD amplitudes with 2 initial spacelike
  legs via generalised BCFW recursion},'' {\em JHEP}, vol.~02, p.~009, 2017.

\bibitem{LipatovEL1}
L.~N. Lipatov, ``{Gauge invariant effective action for high-energy processes in
  QCD},'' {\em Nucl. Phys.}, vol.~B452, pp.~369--400, 1995.

\bibitem{LipatovEL2}
L.~N. Lipatov, ``{Small x physics in perturbative QCD},'' {\em Phys. Rept.},
  vol.~286, pp.~131--198, 1997.

\bibitem{KirschnerEL1}
R.~Kirschner, L.~N. Lipatov, and L.~Szymanowski, ``{Effective action for multi
  - Regge processes in QCD},'' {\em Nucl. Phys.}, vol.~B425, pp.~579--594,
  1994.

\bibitem{KirschnerEL2}
R.~Kirschner, L.~N. Lipatov, and L.~Szymanowski, ``{Symmetry properties of the
  effective action for high-energy scattering in QCD},'' {\em Phys. Rev.},
  vol.~D51, pp.~838--855, 1995.

\bibitem{vanHamerenWL1}
A.~van Hameren, P.~Kotko, and K.~Kutak, ``{Multi-gluon helicity amplitudes with
  one off-shell leg within high energy factorization},'' {\em JHEP}, vol.~12,
  p.~029, 2012.

\bibitem{vanHamerenWL2}
A.~van Hameren, P.~Kotko, and K.~Kutak, ``{Helicity amplitudes for high-energy
  scattering},'' {\em JHEP}, vol.~01, p.~078, 2013.

\bibitem{vanHameren:2017txa}
A.~van Hameren, ``{One-loop amplitudes with an off-shell gluon},'' in {\em
  {17th conference on Elastic and Diffractive Scattering (EDS 17) Prague, Czech
  Republic, June 26-30, 2017}}, 2017.

\bibitem{vanHameren:2017hxx}
A.~van Hameren, ``{Calculating off-shell one-loop amplitudes for
  $k_T$-dependent factorization: a proof of concept},'' 2017.

\bibitem{GribovLevinRyskin}
L.~V. Gribov, E.~M. Levin, and M.~G. Ryskin, ``{Semihard Processes in QCD},''
  {\em Phys. Rept.}, vol.~100, pp.~1--150, 1983.

\bibitem{CataniCiafaloniHautmann}
S.~Catani, M.~Ciafaloni, and F.~Hautmann, ``{High-energy factorization and
  small x heavy flavor production},'' {\em Nucl. Phys.}, vol.~B366,
  pp.~135--188, 1991.

\bibitem{CollinsEllis}
J.~C. Collins and R.~K. Ellis, ``{Heavy quark production in very high-energy
  hadron collisions},'' {\em Nucl. Phys.}, vol.~B360, pp.~3--30, 1991.

\bibitem{CataniHautmann}
S.~Catani and F.~Hautmann, ``{High-energy factorization and small x deep
  inelastic scattering beyond leading order},'' {\em Nucl. Phys.}, vol.~B427,
  pp.~475--524, 1994.

\bibitem{GribovReggeonDiagrams}
V.~N. Gribov, ``{A Reggeon diagram technique},'' {\em Sov. Phys. JETP},
  vol.~26, pp.~414--422, 1968.
\newblock [Zh. Eksp. Teor. Fiz.53,654(1967)].

\bibitem{BFKL1}
L.~N. Lipatov, ``{Reggeization of the Vector Meson and the Vacuum Singularity
  in Nonabelian Gauge Theories},'' {\em Sov. J. Nucl. Phys.}, vol.~23,
  pp.~338--345, 1976.
\newblock [Yad. Fiz.23,642(1976)].

\bibitem{BFKL2}
E.~A. Kuraev, L.~N. Lipatov, and V.~S. Fadin, ``{Multi - Reggeon Processes in
  the Yang-Mills Theory},'' {\em Sov. Phys. JETP}, vol.~44, pp.~443--450, 1976.
\newblock [Zh. Eksp. Teor. Fiz.71,840(1976)].

\bibitem{BFKL3}
V.~S. Fadin, E.~A. Kuraev, and L.~N. Lipatov, ``{On the Pomeranchuk Singularity
  in Asymptotically Free Theories},'' {\em Phys. Lett.}, vol.~B60, pp.~50--52,
  1975.

\bibitem{BFKL4}
E.~A. Kuraev, L.~N. Lipatov, and V.~S. Fadin, ``{The Pomeranchuk Singularity in
  Nonabelian Gauge Theories},'' {\em Sov. Phys. JETP}, vol.~45, pp.~199--204,
  1977.
\newblock [Zh. Eksp. Teor. Fiz.72,377(1977)].

\bibitem{BFKL5}
I.~I. Balitsky and L.~N. Lipatov, ``{The Pomeranchuk Singularity in Quantum
  Chromodynamics},'' {\em Sov. J. Nucl. Phys.}, vol.~28, pp.~822--829, 1978.
\newblock [Yad. Fiz.28,1597(1978)].

\bibitem{NLOBFKL1}
V.~S. Fadin and L.~N. Lipatov, ``{BFKL pomeron in the next-to-leading
  approximation},'' {\em Phys. Lett.}, vol.~B429, pp.~127--134, 1998.

\bibitem{NLOBFKL2}
M.~Ciafaloni and G.~Camici, ``{Energy scale(s) and next-to-leading BFKL
  equation},'' {\em Phys. Lett.}, vol.~B430, pp.~349--354, 1998.

\bibitem{NLOreggeization}
V.~S. Fadin, R.~Fiore, M.~G. Kozlov, and A.~V. Reznichenko, ``{Proof of the
  multi-Regge form of QCD amplitudes with gluon exchanges in the NLA},'' {\em
  Phys. Lett.}, vol.~B639, pp.~74--81, 2006.

\bibitem{BalitskyLectures}
I.~Balitsky, ``{High-energy QCD and Wilson lines},'' 2001.

\bibitem{DixonReview}
L.~J. Dixon, ``{Calculating scattering amplitudes efficiently},'' in {\em {QCD
  and beyond. Proceedings, Theoretical Advanced Study Institute in Elementary
  Particle Physics, TASI-95, Boulder, USA, June 4-30, 1995}}, pp.~539--584,
  1996.

\bibitem{DrummondSuperconformalSymmetry}
J.~M. Drummond, J.~Henn, G.~P. Korchemsky, and E.~Sokatchev, ``{Dual
  superconformal symmetry of scattering amplitudes in N=4 super-Yang-Mills
  theory},'' {\em Nucl. Phys.}, vol.~B828, pp.~317--374, 2010.

\bibitem{Nair}
V.~P. Nair, ``{A Current Algebra for Some Gauge Theory Amplitudes},'' {\em
  Phys. Lett.}, vol.~B214, pp.~215--218, 1988.

\bibitem{ambitwistorReggeons}
L.~V. Bork and A.~I. Onishchenko, ``{Ambitwistor strings and reggeon amplitudes
  in N=4 SYM},'' {\em Phys. Lett.}, vol.~B774, pp.~403--410, 2017.

\bibitem{ambitwistorString4d}
Y.~Geyer, A.~E. Lipstein, and L.~J. Mason, ``{Ambitwistor Strings in Four
  Dimensions},'' {\em Phys. Rev. Lett.}, vol.~113, no.~8, p.~081602, 2014.

\bibitem{DualitySMatrix}
N.~Arkani-Hamed, F.~Cachazo, C.~Cheung, and J.~Kaplan, ``{A Duality For The S
  Matrix},'' {\em JHEP}, vol.~03, p.~020, 2010.

\bibitem{AmplitudesPositiveGrassmannian}
N.~Arkani-Hamed, J.~L. Bourjaily, F.~Cachazo, A.~B. Goncharov, A.~Postnikov,
  and J.~Trnka, {\em {Scattering Amplitudes and the Positive Grassmannian}}.
\newblock Cambridge University Press, 2012.

\bibitem{BCFW1}
R.~Britto, F.~Cachazo, and B.~Feng, ``{New recursion relations for tree
  amplitudes of gluons},'' {\em Nucl. Phys.}, vol.~B715, pp.~499--522, 2005.

\bibitem{BCFW2}
R.~Britto, F.~Cachazo, B.~Feng, and E.~Witten, ``{Direct proof of tree-level
  recursion relation in Yang-Mills theory},'' {\em Phys. Rev. Lett.}, vol.~94,
  p.~181602, 2005.

\bibitem{AllLoopIntegrandN4SYM}
N.~Arkani-Hamed, J.~L. Bourjaily, F.~Cachazo, S.~Caron-Huot, and J.~Trnka,
  ``{The All-Loop Integrand For Scattering Amplitudes in Planar N=4 SYM},''
  {\em JHEP}, vol.~01, p.~041, 2011.

\bibitem{MomentumTwistors}
A.~Hodges, ``{Eliminating spurious poles from gauge-theoretic amplitudes},''
  {\em JHEP}, vol.~05, p.~135, 2013.

\bibitem{LocalPhysicsGrassmanian}
N.~Arkani-Hamed, J.~Bourjaily, F.~Cachazo, and J.~Trnka, ``{Local Spacetime
  Physics from the Grassmannian},'' {\em JHEP}, vol.~01, p.~108, 2011.

\bibitem{BCFWmathematica}
J.~L. Bourjaily, ``{Efficient Tree-Amplitudes in N=4: Automatic BCFW Recursion
  in Mathematica},'' 2010.

\bibitem{AmplituhedronFromMomentumTwistors}
Y.~Bai and S.~He, ``{The Amplituhedron from Momentum Twistor Diagrams},'' {\em
  JHEP}, vol.~02, p.~065, 2015.

\bibitem{Boels:2010nw}
R.~H. Boels, ``{On BCFW shifts of integrands and integrals},'' {\em JHEP},
  vol.~11, p.~113, 2010.

\bibitem{Boels:2016jmi}
R.~H. Boels and H.~Luo, ``{On-shell recursion relations for generic
  integrands},'' 2016.

\bibitem{LocalIntegrands}
N.~Arkani-Hamed, J.~L. Bourjaily, F.~Cachazo, and J.~Trnka, ``{Local Integrals
  for Planar Scattering Amplitudes},'' {\em JHEP}, vol.~06, p.~125, 2012.

\bibitem{NLOBFKL-QCD-SUSY}
A.~V. Kotikov and L.~N. Lipatov, ``{NLO corrections to the BFKL equation in QCD
  and in supersymmetric gauge theories},'' {\em Nucl. Phys.}, vol.~B582,
  pp.~19--43, 2000.

\bibitem{DGLAP-BFKL-N4SYM1}
A.~V. Kotikov and L.~N. Lipatov, ``{DGLAP and BFKL evolution equations in the
  N=4 supersymmetric gauge theory},'' in {\em {35th Annual Winter School on
  Nuclear and Particle Physics Repino, Russia, February 19-25, 2001}}, 2001.

\bibitem{DGLAP-BFKL-N4SYM2}
A.~V. Kotikov and L.~N. Lipatov, ``{DGLAP and BFKL equations in the $N=4$
  supersymmetric gauge theory},'' {\em Nucl. Phys.}, vol.~B661, pp.~19--61,
  2003.
\newblock [Erratum: Nucl. Phys.B685,405(2004)].

\bibitem{ConformalProperties4point}
J.~M. Drummond, G.~P. Korchemsky, and E.~Sokatchev, ``{Conformal properties of
  four-gluon planar amplitudes and Wilson loops},'' {\em Nucl. Phys.},
  vol.~B795, pp.~385--408, 2008.

\bibitem{Du:2014jwa}
P.~Du, G.~Chen, and Y.-K.~E. Cheung, ``{Permutation relations of generalized
  Yangian Invariants, unitarity cuts, and scattering amplitudes},'' {\em JHEP},
  vol.~09, p.~115, 2014.

\end{thebibliography}

\end{document}